%% file: EXO-22-020_temp.tex
\pdfoutput=1
\documentclass[11pt,twoside,a4paper,cmspaper,final,collab]{cms-tdr}
\def\svnVersion{3ec501c}\def\svnDate{2024/06/06}\def\cmsCernNoTag{CERN-EP-2024-031}\def\cmsCernDate{\today}\def\cmsMessage{Published in Physical Review D as \href{http://dx.doi.org/10.1103/PhysRevD.109.112005}{\doi{10.1103/PhysRevD.109.112005}.}}
\begin{document}\cmsNoteHeader{EXO-22-020}

\newcommand{\dxy}{\ensuremath{d_{\mathrm{xy}}}\xspace}
\newcommand{\dbv}{\ensuremath{d_{\mathrm{BV}}}\xspace}
\newcommand{\dbverr}{\ensuremath{\sigma_{\dbv}}\xspace}
\newcommand{\ntk}{\ensuremath{n_{\text{track}}}\xspace}
\newcommand{\mlscore}{\ensuremath{S_{\text{ML}}}\xspace}
\newcommand{\dvv}{\ensuremath{d_{\mathrm{vv}}}\xspace}
\newcommand{\dtv}{\ensuremath{d_{\mathrm{tv}}}\xspace}
\newcommand{\dvverr}{\ensuremath{\sigma_{\dvv}}\xspace}
\newcommand{\dtverr}{\ensuremath{\sigma_{\dtv}}\xspace}
\newcommand{\dvvsig}{\ensuremath{\abs{\dvv/\dvverr}}\xspace}
\newcommand{\dtvsig}{\ensuremath{\abs{\dtv/\dtverr}}\xspace}
\newcommand{\Rr}{\ensuremath{R_{\text{r}}}\xspace}
\newcommand{\Rs}{\ensuremath{R_{\text{s}}}\xspace}
\newcommand{\phiR}{\ensuremath{\phi_{R}}\xspace}
\newcommand{\phiO}{\ensuremath{\phi_{O}}\xspace}
\newcommand{\phiout}{\ensuremath{\phi_{\text{output}}}\xspace}
\newcommand{\lbce}{\ensuremath{L_{\text{bce}}}\xspace}
\newcommand{\lreg}{\ensuremath{L_{\text{reg}}}\xspace}
\newcommand{\ldcorr}{\ensuremath{L_{\text{dcorr}}}\xspace}
\newcommand{\lamreg}{\ensuremath{\lambda_{\text{reg}}}\xspace}
\newcommand{\lamdcorr}{\ensuremath{\lambda_{\text{dcorr}}}\xspace}
\newcommand{\mSUSY}{\ensuremath{m_{\text{SUSY}}}\xspace}
\newcommand{\mPSg}{\ensuremath{m_{\PSg}}\xspace}
\ifthenelse{\boolean{cms@external}}{\providecommand{\NA}{\ensuremath{\cdots}\xspace}}{\providecommand{\NA}{\ensuremath{\text{---}}\xspace}}
\newlength\cmsTabSkip\setlength{\cmsTabSkip}{1ex}
\ifthenelse{\boolean{cms@external}}{\providecommand{\cmsLeft}{upper\xspace}}{\providecommand{\cmsLeft}{left\xspace}}
\ifthenelse{\boolean{cms@external}}{\providecommand{\cmsRight}{lower\xspace}}{\providecommand{\cmsRight}{right\xspace}}
\ifthenelse{\boolean{cms@external}}{\providecommand{\cmsLLeft}{Upper\xspace}}{\providecommand{\cmsLLeft}{Left\xspace}}
\ifthenelse{\boolean{cms@external}}{\providecommand{\cmsRRight}{Lower\xspace}}{\providecommand{\cmsRRight}{Right\xspace}}

\cmsNoteHeader{EXO-22-020}
\title{Search for long-lived particles using displaced vertices and missing transverse momentum in proton-proton collisions at \texorpdfstring{$\sqrt{s}=13\TeV$}{sqrt(s) = 13 TeV}}

\date{\today}

\abstract{
A search for the production of long-lived particles in proton-proton collisions at a center-of-mass energy of 13\TeV at the CERN LHC is presented. The search is based on data collected by the CMS experiment in 2016--2018, corresponding to a total integrated luminosity of 137\fbinv. This search is designed to be sensitive to long-lived particles with mean proper decay lengths between 0.1 and 1000\mm, whose decay products produce a final state with at least one displaced vertex and missing transverse momentum. A machine learning algorithm, which improves the background rejection power by more than an order of magnitude, is applied to improve the sensitivity. The observation is consistent with the standard model background prediction, and the results are used to constrain split supersymmetry (SUSY) and gauge-mediated SUSY breaking models with different gluino mean proper decay lengths and masses. This search is the first CMS search that shows sensitivity to hadronically decaying long-lived particles from signals with mass differences between the gluino and neutralino below 100\GeV. It sets the most stringent limits to date for split-SUSY models and gauge-mediated SUSY breaking models with gluino proper decay length less than 6\mm.
}

\hypersetup{
pdfauthor={CMS Collaboration},
pdftitle={Search for long-lived particles using displaced vertices and missing transverse momentum in proton-proton collisions at sqrt(s)=13 TeV},
pdfsubject={CMS},
pdfkeywords={CMS, beyond standard model, long-lived particles, LLPs}} 

\maketitle 
\section{Introduction}

{\tolerance=800 
Many theoretical models beyond the standard model (SM) predict the existence of long-lived particles (LLPs) with measurable decay lengths.
Examples include split supersymmetry (SUSY)~\cite{Giudice:2004tc, Hewett:2004nw,ArkaniHamed:2004yi,Gambino:2005eh,Arvanitaki:2012ps,ArkaniHamed:2012gw}, SUSY with weak $R$-parity violation~\cite{Fayet:1974pd,Farrar:1978xj,Weinberg:1981wj,Barbier:2004ez}, gauge-mediated SUSY breaking (GMSB)~\cite{GIUDICE1999419,Meade:2008wd,Buican:2008ws}, stealth SUSY~\cite{Fan:2011yu,Fan:2012jf}, hidden-valley models~\cite{Strassler:2006im,Strassler:2006ri,Han:2007ae}, 
and twin Higgs models~\cite{Chacko:2005pe,Curtin:2015fna,Cheng:2015buv}. New particles predicted by these models can be produced at the CERN LHC, and LLP searches provide an opportunity to discover particles that have not been excluded by previous searches.
\par}

The search presented here is sensitive to LLPs whose decays include multiple charged particles and one weakly interacting neutral particle. To target this signature, the search is designed to be sensitive to topologies that include missing transverse momentum (\ptmiss) and at least one displaced vertex. A displaced vertex is reconstructed by identifying charged-particle trajectories converging at a point in space that is at a measurable distance from the proton-proton ($\Pp\Pp$) interaction point. Reconstructed vertices are required to be within a transverse displacement of approximately 2\cm, which ensures they are composed of well-measured trajectories and eliminates the possibility of background vertices caused by SM particle interactions with the CMS beam pipe.

The ATLAS and CMS Collaborations have previously performed searches for LLPs in signatures with displaced vertices~\cite{ATLAS:2017tny, ATLAS:2020xyo, EXO-19-013, CMS:2020iwv, ATLAS:2023oti}. Those searches target LLPs that either decay at least 1\mm away from the interaction point, or have most of their energy decay into SM particles. The search presented in this paper is designed to cover the gaps left by previous searches, which include a wider range of LLP mean proper decay lengths ($c\tau$), from 0.1 to 1000\mm, and LLPs in ``compressed'' scenarios, where most of the LLP energy is carried away by undetected weakly interacting particles.

This search includes several improvements over the previous CMS displaced vertex search~\cite{EXO-19-013}. One such improvement is the inclusion of events with as few as one displaced vertex. This allows sensitivity to new physics models with only one LLP in the final state. It also enhances sensitivity to beyond the SM scenarios in which the reconstruction of displaced vertices is challenging, such as models involving lower energy carried by charged particles in the final states or cases where only one of the LLP decays occurs within the detector. Although requiring only one displaced vertex increases the signal efficiency, it also includes more background events. To further discriminate signal and background events, the ``interaction network'' (IN)~\cite{Moreno:2019bmu,Moreno:2019neq}, a machine learning (ML) algorithm, is introduced to improve the sensitivity to the target signature. An alternative selection without using ML was studied and found to have a factor of 8 more background events for smaller signal efficiency. In addition, this search includes events with an overall total transverse momentum (\pt) imbalance, making it sensitive to final states with a significantly smaller scalar sum of jet \pt. The search uses events collected in $\Pp\Pp$ collisions at the LHC in 2016--2018, corresponding to an integrated luminosity of 137\fbinv.

\begin{figure}[hbtp]
\centering
\includegraphics[width=0.49\textwidth]{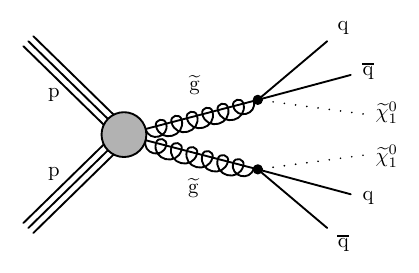}
\includegraphics[width=0.49\textwidth]{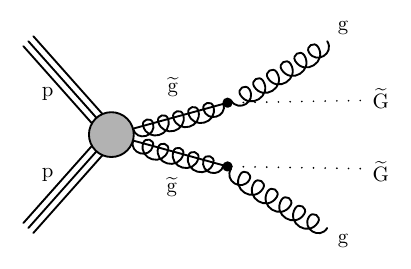}
\caption{Diagrams of the split-SUSY model (\cmsLeft) and GMSB SUSY model (\cmsRight). In the split-SUSY model, a pair of long-lived gluinos is produced, and each decays to two quarks and one neutralino. In the GMSB SUSY model, a pair of long-lived gluinos is produced, and each decays to a gluon and a gravitino.
}
\label{fig:signal_diagram}
\end{figure}

This analysis uses split SUSY and GMSB SUSY as benchmark signal models, as shown in Fig.~\ref{fig:signal_diagram}. The split-SUSY model features pair-produced, long-lived gluinos (\PSg) that each decay into two quarks (\PQq) and a neutralino (\PSGc). In this model, the SUSY breaking scale (\mSUSY) is assumed to be $\mSUSY \gg 10^{6}\TeV$~\cite{Hewett:2004nw}, and all scalar masses are set to that scale, except for a single, fine-tuned, Higgs boson mass. The \PSg decays through a high-mass, virtual squark resulting in a long \PSg lifetime. 
In the GMSB SUSY model, \PSg are pair-produced and decay to a gluon and a nearly massless gravitino (\sGra). The decay is suppressed by \mSUSY, and thus the \PSg is long-lived. The lifetime of the gluino in both models is given by
\begin{linenomath}
\begin{equation}
\tau \simeq 8\,\Big(\frac{\mSUSY}{10^{6}\TeV}\Big)^{4}\,\Big(\frac{1\TeV}{\mPSg}\Big)^{5}\unit{s},
\end{equation}
\end{linenomath}
where \mPSg corresponds to the \PSg mass~\cite{Hewett:2004nw,GIUDICE1999419}.
Previously, the CMS Collaboration performed a search for the split-SUSY signal model~\cite{CMS:2018ncl}; however, the displaced signature was not targeted and the resulting large number of background events decreased the sensitivity.

The paper is organized as follows. Section~\ref{sec:detector} presents an overview of the CMS detector. Section~\ref{sec:datasets} summarizes the data and simulation samples used in this search. We describe the displaced vertex reconstruction algorithm in Section~\ref{sec:vtxreco}, followed by the discussion of the ML algorithm in Section~\ref{sec:ml}. The event selection criteria and the determination of vertex reconstruction and ML tagging efficiency are summarized in Section~\ref{sec:vtx_ml_eff}. In Section~\ref{sec:bkgest}, we explain the background estimation procedure based on control samples in data. Section~\ref{sec:syst} discusses systematic uncertainties in signal efficiency and background estimation. Section~\ref{sec:results} presents the results and statistical interpretation. Finally, the paper is summarized in Section~\ref{sec:summary}.
Tabulated results are provided in the HEPData record for this analysis~\cite{hepdata}.

\section{The CMS detector}
\label{sec:detector}

The central feature of the CMS apparatus is a superconducting solenoid of 6\unit{m} inner diameter, providing a magnetic field of 3.8\unit{T}. Within the solenoid volume are a silicon pixel and strip tracker, a lead tungstate crystal electromagnetic calorimeter (ECAL), and a brass and scintillator hadron calorimeter (HCAL), each composed of a barrel and two endcap sections. Forward calorimeters extend the pseudorapidity ($\eta$) coverage provided by the barrel and endcap detectors. 
A more detailed description of the CMS detector, together with a definition of the coordinate system used and the relevant kinematic variables, can be found in Ref.~\cite{CMS:2008xjf}.

The silicon tracker used in 2016 measured charged particles within the range $\abs{\eta} < 2.5$. For nonisolated particles within the ranges of $1 < \pt < 10\GeV$ and $\abs{\eta} < 1.4$, the track resolutions were typically 1.5\% in \pt and 25--90\mum in the transverse impact parameter (\dxy)~\cite{CMS:2014pgm}. At the start of 2017, a new pixel detector was installed~\cite{Phase1Pixel}; the upgraded tracker measured particles up to $\abs{\eta} < 3.0$ with typical resolutions of 1.5\% in \pt and 20--75\mum in \dxy~\cite{DP-2020-049} for nonisolated particles of $1 < \pt < 10\GeV$.

A particle-flow (PF) algorithm~\cite{CMS:2017yfk} aims to reconstruct and identify each individual particle in an event, with an optimized combination of information from the various elements of the CMS detector. The energy of photons is obtained from the ECAL measurement. The energy of electrons is determined from a combination of the electron momentum at the primary interaction vertex as determined by the tracker, the energy of the corresponding ECAL cluster, and the energy sum of all bremsstrahlung photons spatially compatible with originating from the electron track. The energy of muons is obtained from the curvature of the corresponding track. The energy of charged hadrons is determined from a combination of their momentum measured in the tracker and the matching ECAL and HCAL energy deposits, corrected for the response function of the calorimeters to hadronic showers. Finally, the energy of neutral hadrons is obtained from the corresponding corrected ECAL and HCAL energies.

The primary vertex is taken to be the vertex corresponding to the hardest scattering in the event, evaluated using tracking information alone, as described in Section 9.4.1 of Ref.~\cite{CMS-TDR-15-02}.

For each event, hadronic jets are clustered from PF candidates using the infrared- and collinear-safe anti-\kt algorithm~\cite{Cacciari:2008gp, Cacciari:2011ma} with a distance parameter of 0.4. Jet momentum is determined as the vectorial sum of all particle momenta in the jet, and is found from simulation to be, on average, within 5 to 10\% of the true momentum over the entire \pt spectrum and detector acceptance. Additional $\Pp\Pp$ interactions within the same or nearby bunch crossings (referred to as pileup) can contribute additional tracks and calorimetric energy depositions to the jet momentum. To mitigate this effect, charged particles identified to be originating from pileup vertices are discarded and an offset correction is applied to correct for remaining contributions. Jet energy corrections are derived from simulation to bring the measured response of jets to that of particle level jets on average. In situ measurements of the momentum balance in dijet, $\text{photon}{+}\text{jet}$, $\PZ{+}\text{jet}$, and multijet events are used to account for any residual differences in the jet energy scale between data and simulation~\cite{CMS:2016lmd}. The jet energy resolution amounts typically to 15--20\% at 30\GeV, 10\% at 100\GeV, and 5\% at 1\TeV~\cite{CMS:2016lmd}. Additional selection criteria are applied to each jet to remove jets potentially dominated by anomalous contributions from various subdetector components or reconstruction failures~\cite{CMS:2017wyc}.

The \ptvecmiss is computed as the negative vector \pt sum of all the PF candidates except muons in an event, and its magnitude is denoted as \ptmiss~\cite{CMS:2019ctu}. The \ptvecmiss is modified to account for corrections to the energy scale of the reconstructed jets in the event. Anomalous events with high \ptmiss can be due to a variety of reconstruction failures, detector malfunctions, or noncollision backgrounds. Such events are rejected by event filters that are designed to identify more than 85--90\% of the spurious high-\ptmiss events with a mistagging rate less than 0.1\%~\cite{CMS:2019ctu}.

Events of interest are selected using a two-tiered trigger system. The first level, referred to as the level-1 (L1) trigger, is composed of custom hardware processors and uses information from the calorimeters and muon detectors to select events at a rate of around 100\unit{kHz} within a fixed latency of about 4\mus~\cite{CMS:2020cmk}. The second level, known as the high-level trigger, consists of a farm of processors running a version of the full event reconstruction software optimized for fast processing, and reduces the event rate to around 1\unit{kHz} before data storage~\cite{CMS:2016ngn}.

\section{Event samples and preselection}
\label{sec:datasets}

Events in both data and simulation are selected using a trigger requiring $\ptmiss > 120\GeV$. Events are further required to pass a set of quality filters~\cite{CMS:2019ctu} to reduce the possibility of mismeasuring \ptmiss and to have a well-reconstructed primary vertex. In addition, an offline requirement of $\ptmiss > 200\GeV$ is applied to ensure consistent trigger efficiencies in data and simulation. These selections are referred to as event preselection criteria. Since the CMS detector operated under different conditions during successive data-taking periods, events collected in 2016, 2017, and 2018 are analyzed separately.

{\tolerance=800 
Signal events are simulated using \PYTHIA 8.240~\cite{Sjostrand:2014zea} with the CP2 tune~\cite{CMS:2019csb} and the NNPDF3.1 leading-order (LO)~\cite{NNPDF:2017mvq} set of parton distribution functions (PDFs). The gluino production cross sections are then normalized to next-to-next-to-LO (NNLO) in quantum chromodynamics (QCD) and next-to-next-to-leading-logarithm (NNLL) soft-gluon resummation precision~\cite{Beenakker:2016lwe}. In order to account for possible kinematic effects of bound state formation, production of bound states that include a gluino is modeled with \PYTHIA. The probability of forming a bound state of gluinos alone is set to the default value of 0.1. Since the search targets only displaced vertices within the beam pipe, material interactions with LLPs and their potential bound states are not considered. The split-SUSY samples are generated for various \mPSg hypotheses in the range from 1200 to 2600\GeV, with the mass splitting between the \PSg and \PSGc ranging from 20 to 800\GeV and the gluino $c\tau$ ranging from 0.1 to 1000\mm. The GMSB SUSY samples are generated with \mPSg ranging from 1800 to 2800\GeV and the gluino $c\tau$ ranging from 0.1 to 1000\mm.
\par}

The dominant background processes include jets produced through the strong interaction, referred to as QCD events, and \ttbar. Other contributions arise from \PW and \PZ boson production in association with jets ($\PW/\PZ+\text{jets}$), single top quark production, and diboson production ($\PW\PW$, $\PW\PZ$, and $\PZ\PZ$). The QCD and $\PW/\PZ+\text{jets}$ events are generated using \MGvATNLO 2.6.5~\cite{Alwall:2014hca} at LO in QCD, with the MLM~\cite{Alwall:2007fs} prescription of matching jets from matrix element calculations to those from parton showers. The \ttbar events are generated using \MGvATNLO 2.6.1~\cite{Alwall:2014hca} at next-to-LO (NLO), with the FxFx~\cite{Frederix:2012ps} prescription of jet matching. Single top quark production is generated using \POWHEG~\cite{Nason:2004rx, Frixione:2007vw, Alioli:2010xd, Re:2010bp, Alioli:2009je} (version 2.0 for $t$-channel and 1.0 for $\PQt\PW$-channel production) at NLO in QCD and the diboson processes are generated using \PYTHIA 8.240. The NNPDF3.1 NNLO~\cite{NNPDF:2014otw} PDF set is used for all background processes. The background event generators are interfaced with \PYTHIA to model the parton showering and fragmentation. The \PYTHIA parameters affecting the description of the underlying event are set to the CP5 tune~\cite{CMS:2019csb}.

{\tolerance=800 
All generated events are processed through a simulation of the CMS detector based on \GEANTfour~\cite{GEANT4:2002zbu}, and are reconstructed with the same algorithms as used for data. Simulated minimum bias events are superimposed on the hard interaction to describe the effect of pileup, and samples are weighted to match the pileup distribution observed in data. 
\par}

\section{Vertex reconstruction}
\label{sec:vtxreco}

A detailed description of the displaced vertex reconstruction algorithm can be found in Ref.~\cite{EXO-19-013}. The displaced vertices are constructed from tracks. A set of quality criteria is applied to remove tracks with large reconstruction uncertainty, and a displacement criterion is applied to reduce the contribution of tracks from prompt SM processes. The quality requirements include $\pt > 1\GeV$, at least one associated detector hit measured by the innermost layer of the pixel detector, at least two associated detector hits measured by the pixel detector, and at least six associated detector hits measured by the silicon strip detector. The displacement criterion used is based on \dxy measured with respect to the beam axis and its uncertainty ($\sigma_{\dxy}$). In this search, the tracks selected to construct vertices are required to have $\abs{\dxy/\sigma_{\dxy}}>4$. 

Each pair of selected tracks is used to construct a vertex using the Kalman filtering approach~\cite{Fruhwirth:1987fm, Billoir:1990we, Billoir:1990vz}, and the vertex is considered valid if its $\chi^2$ per degree of freedom is less than five. 
These reconstructed vertices then undergo an iterative merging procedure. This procedure utilizes the significance, defined as the ratio of a quantity and its uncertainty, of the 3D distance between vertices (\dvv and \dvverr), \dvvsig, and the significance of the 3D impact parameter between tracks and vertices (\dtv and \dtverr), \dtvsig. First, for each pair of vertices that share at least one track and have $\dvvsig < 4$, a new vertex is fit by the Kalman filtering method using all tracks associated with the pair of vertices. If the resulting vertex satisfies the $\chi^2$ requirement, the pair of vertices is replaced by a single merged vertex. If not, the vertices are not merged and the shared track is assigned to one or neither of the two vertices according to the following arbitration criteria. If \dtvsig formed from either of the vertices is greater than five, the track is dropped from that vertex; if \dtvsig formed from both vertices is less than 1.5, the track is assigned to the vertex with the most tracks; otherwise, the track is assigned to the vertex that has the smaller \dtvsig. The resulting vertices then replace the original vertices if they meet the $\chi^2$ requirement. This process is repeated until no vertices share tracks.

The final step of the vertex reconstruction aims to suppress ones in which tracks from different primary vertices coincidentally overlap. For each displaced vertex, we repeat the vertex fit with one track removed at a time and calculate the resulting change along the beam axis. If the vertex position along the beam axis shifts by more than 50\mum, which was chosen to maximize signal significance in Ref.~\cite{EXO-19-013}, the track is removed from the vertex, and the procedure is iteratively repeated with the remaining tracks. The procedure stops if there are only two tracks in the vertex. All resulting vertices are required to satisfy the $\chi^2$ requirement.

A set of selection criteria, which is similar to that of Ref.~\cite{EXO-19-013}, is applied to the vertices in order to suppress vertices that arise from SM processes. First, the number of tracks in each vertex, \ntk, must be $\geq$3. To suppress vertices from material interactions, the transverse distance from the center of the beam pipe to the vertex must be less than the radius of the beam pipe, which is approximately 2\cm. Finally, the uncertainty in the transverse distance between the beam axis and the vertex (\dbv), \dbverr, must be $<$25\mum. This criterion helps to reject vertices from decays of Lorentz-boosted particles, such as \PB mesons, which tend to have larger \dbverr.

This search specifically targets events with at least one displaced vertex. The value of \ntk serves as a discriminant between signal and background events. Vertices with \ntk of at least five are considered as signal vertices, while vertices with \ntk of three or four are used as control samples for background estimation. In cases where there are multiple displaced vertices, the one with the highest number of tracks is selected for signal extraction.

\section{Machine learning algorithm}
\label{sec:ml}

The vertex reconstruction and selection largely suppress backgrounds from SM LLPs, such as \PB and \PK mesons. The remaining background events are dominated by \ttbar and QCD events, with background vertices originating from unrelated tracks that coincidentally overlap with each other. To further reduce such background vertices, the IN algorithm is used in this analysis. This section introduces the architecture, training and testing, and performance evaluation of the IN. 

\subsection{Interaction network description}

An IN is a type of graph neural network~\cite{gnn} that takes graphs, a data structure that is composed of nodes and edges, as input. Nodes can be used to represent objects and edges can be used to represent relations between different nodes. 
With all of these features, INs are capable of calculating the interactions between different objects in a multibody system and thus predict the final states of all physical objects in the system. We apply an IN to the analysis of CMS events by treating each collision as a multibody interaction. The unique topology of LLP decays produces relationships between tracks that an IN can be trained to identify. In this analysis, we train an IN to calculate the relationships between tracks and learn that tracks resulting from LLP decays originate from displaced spatial points. As a result, the IN can output a discriminant value for assessing whether a given event contains LLP decays predicted in scenarios beyond the SM.

Given that variables associated with the vertex reconstruction are very powerful at discriminating signal and background, including those variables in the IN does not improve the discrimination of background events as the IN will mostly rely on the reconstructed vertices. To enable the IN to exploit as much additional information as possible, the IN is provided tracking information but no information about the reconstructed vertices themselves. This approach allows the IN to make use of the topologies of tracks in signal and background events without the help of the existing vertex reconstruction algorithm.
The output of the IN is used together with variables from reconstructed vertices to improve the search sensitivity, as described in Section~\ref{sec:vtx_ml_eff}. 

Since the IN is computationally intensive, for each event, only the first 50 \pt-ordered tracks are used as input. Tracks are required to satisfy the quality criteria but not the displacement criterion described in Section~\ref{sec:vtxreco}, a choice that provides the IN the opportunity to consider additional tracks from LLP decays whose trajectories coincide with the beam axis. For the split-SUSY signal model, the average number of tracks used by the IN that are assigned to reconstructed signal vertices typically ranges from 4 to 10, depending on mass splitting and gluino $c\tau$. In the IN, the input tracks are implemented as nodes in the graph while edges are composed of relations between any pairs of tracks. The tracking information used in the IN includes \pt, $\eta$, the azimuthal angle ($\phi$), \dxy and its significance, and the longitudinal impact parameter $d_{\mathrm{z}}$ and its significance.

The IN is implemented using standard deep neural network building blocks, such as multilayer perceptrons (MLPs)~\cite{Murtagh1991} and matrix operations including addition, subtraction, multiplication, and concatenation. The input is constructed from several matrices: the variable matrix $O$, which contains track information; the send relation matrix \Rs, which represents the interaction a track sends to other tracks; and the receive relation matrix \Rr, which represents the interaction a track receives from other tracks. The interaction described in the IN is an abstract concept that could include any relationship between a pair of tracks. 

Both \Rr and \Rs are structured such that each row represents a single track, and each column corresponds to the various interactions that tracks can send or receive. 
We want the IN to consider all possible relationships between tracks, so each element of \Rr is set to zero or one such that every track can receive interactions from every track except itself, and each element of \Rs is similarly set to zero or one such that every track can send interactions to every track except itself. 

\begin{figure*}[hbtp]
\centering
\includegraphics[width=0.98\textwidth]{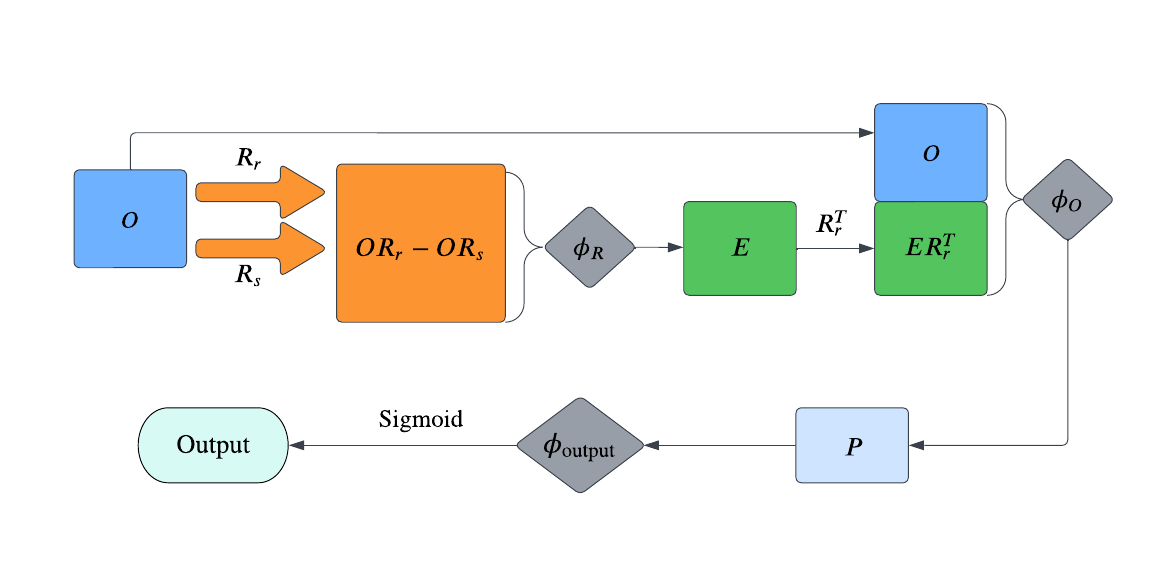}
\caption{
An illustration of the architecture of the IN, where the flow of data is indicated by arrows. Rectangular boxes represent data matrices, while diamonds represent multilayer perceptrons (MLPs). The original input information ($O$) is integrated with relation matrices (\Rr and \Rs) to form a graph that captures interactions between tracks. This graph is subsequently processed by an MLP (\phiR) to compute the effect ($E$) of the interactions. The effect is then combined with \Rr and merged with the original input $O$. To assess the influence ($P$) of the effect on the original information, it undergoes further processing via another MLP (\phiO). Finally, the influence is passed through an MLP (\phiout) and a sigmoid function to produce the final output.
}
\label{fig:ml_diagram}
\end{figure*}

A diagram of the architecture of the IN is shown in Fig.~\ref{fig:ml_diagram}. The IN calculates the state after tracks receive and send relations with other tracks by $O \Rr$ and $O \Rs$, and combines these calculated elements together as a single matrix by $O \Rr - O \Rs$. 
The calculated matrix is passed into a five-layer MLP, labeled \phiR, with 50 nodes per layer. The matrix $E$, which is the output of \phiR, is interpreted as the effect of all relations having acted on each track. The effect on every track is represented as an array of length 20 in matrix $E$. The effect matrix $E$ is multiplied with the transpose of matrix \Rr to calculate the effect received by objects. The result is combined with the original variable matrix $O$ to form a new matrix $C$, which contains the original information together with the effects having acted on tracks. The matrix $C$ is processed by another MLP with two parts. The first part, labeled as \phiO, has two layers with 50 nodes in each layer and outputs matrix $P$. The second part, labeled as \phiout, has three layers with 100 nodes in each layer. It takes matrix $P$ as input and outputs a vector that contains a single MLP output value for each event. Finally, this output value is processed with a sigmoid function to produce an IN output score (\mlscore), which predicts whether an event contains an LLP decay or not.

\subsection{Training and testing}
\label{subsec:ml_train_test}

As a discriminant that predicts whether an event is signal or background, the \mlscore is designed to have a value of 0 corresponding to background events and a value of 1 corresponding to signal events. The training process aims to optimize the discriminating power of the \mlscore. The IN is trained by iteratively minimizing the loss function~\cite{ml_loss_function}, which is composed of several parts: a binary cross entropy \lbce~\cite{ml_loss_function}, an L2 regularization \lreg~\cite{ml_l2reg}, and a distance correlation term \ldcorr~\cite{Kasieczka:2020yyl}.
The binary cross entropy \lbce is a commonly used metric for binary discriminant tasks. It quantifies the discrepancy between the predicted outcomes and the desired outcomes. The L2 regularization term \lreg is calculated as the sum of the squared magnitudes of the model parameters and serves to prevent overfitting by constraining these parameters. 
In background events, a distance correlation (DisCo)~\cite{ABCDisCo} method is used to decorrelate \mlscore from \ntk so that these two variables can be used in the background estimation method described in Section~\ref{sec:bkgest}. The DisCo method adds \ldcorr to the IN loss function, which is based on the degree of correlation between \mlscore and the \ntk value of the vertex with the largest number of tracks in background events. We note that fewer than 0.05\% of background events contain more than one reconstructed displaced vertex. The complete IN loss function is then $L=\lbce+\lamreg \lreg+\lamdcorr \ldcorr$, where \lamreg and \lamdcorr are hyperparameters that control the strength of each restriction. In each iteration, the IN processes all training sample events in batches of 128 events and updates the MLP parameters after processing each batch using the Adam algorithm~\cite{ml_adam} to minimize the loss function. Matrix elements of \Rr and \Rs are not updated during the training since they are just used to establish the relationships between tracks.

The events used for training, validation, and testing are required to have at least one reconstructed vertex with $\ntk \geq 3$ and pass all event preselection requirements except the offline \ptmiss selection. Training and validation events are required to have $80 < \ptmiss < 200\GeV$ to ensure orthogonality with testing events, which have $\ptmiss > 200\GeV$. Within the training and validation events, 85\% of them are used for training, and 15\% are used for validation. To avoid any bias introduced by potential mismodeling of simulated events, 17\,067 data events and 31\,165 simulated background events are mixed together and labeled as background events during the training of the IN. The simulated background events are drawn from different simulated SM processes in proportion to their cross section. To avoid potential bias, data events that include vertices with $\ntk \geq 5$ are excluded from the training and validation samples. Split-SUSY signal samples with different lifetimes and masses are combined and labeled as signal events in the training, and events from different data-taking periods are combined during training. Specifically, 91\,013 events are used for training and 17\,065 events are used for validation.

Hyperparameter values are set to $\lamreg = 0.00005$ and $\lamdcorr = 0.65$, which ensures that the model is not overtrained and \mlscore is uncorrelated with \ntk. A learning rate of 0.0003 is chosen to avoid the case that the training does not converge because of a large learning rate. Models with different combinations of hyperparameters were tested, and the trained model that provides the best discriminating power while satisfying the decorrelation requirement between \mlscore and 
\ntk is used for all events in all data-taking periods.

\subsection{Performance}

\begin{figure*}[hbtp]
\centering
\includegraphics[width=0.42\textwidth]{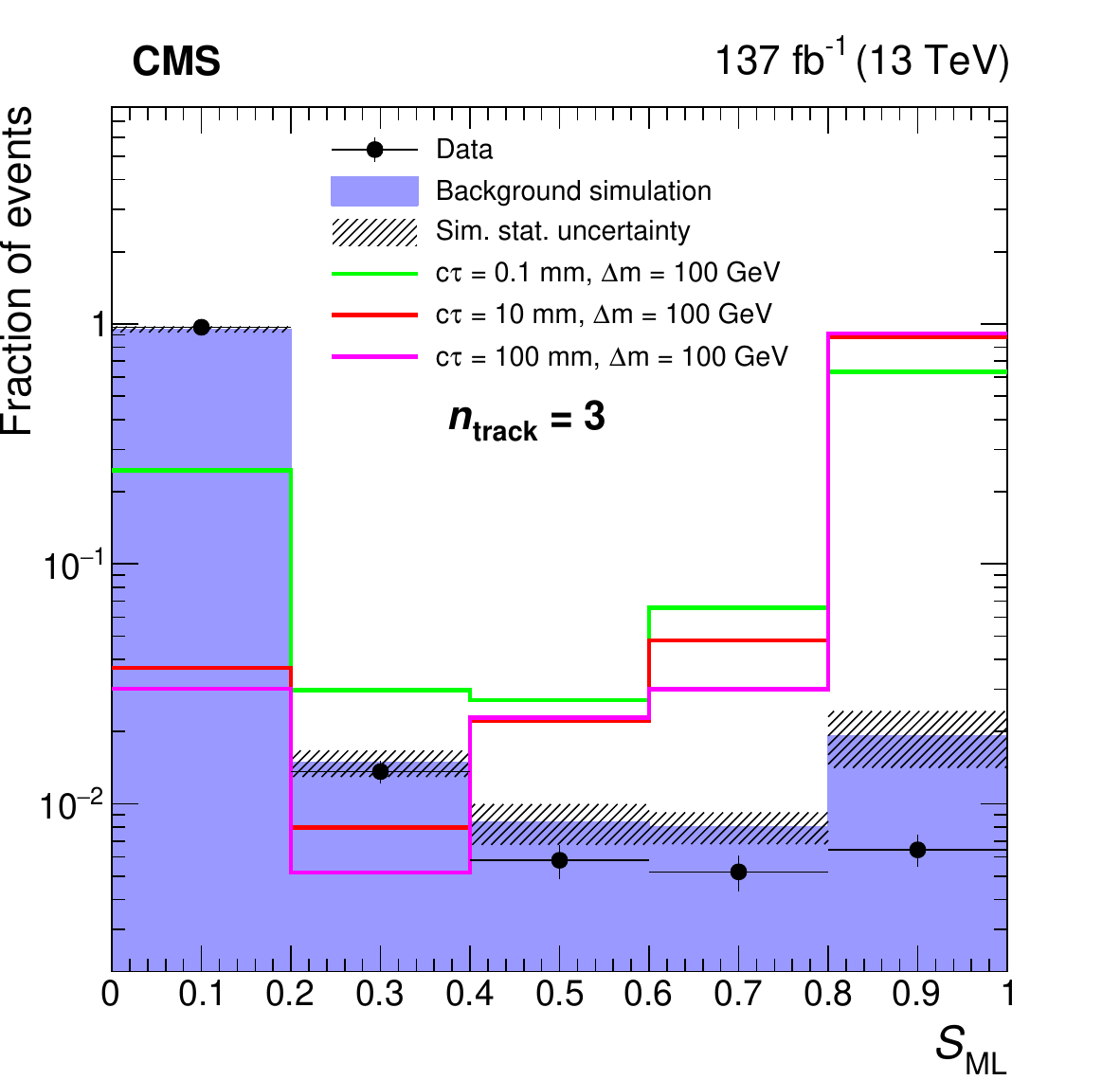}
\includegraphics[width=0.42\textwidth]{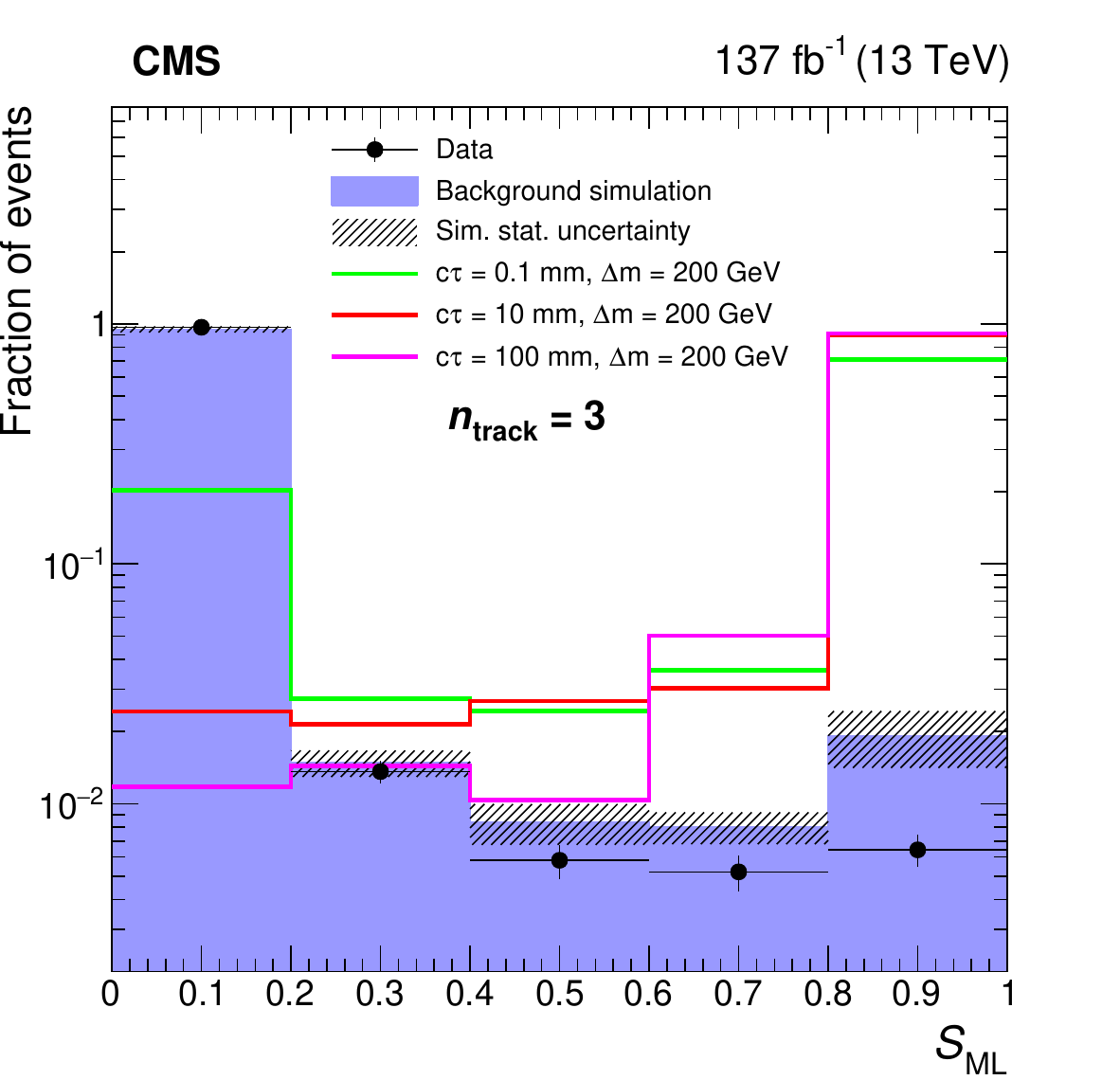}
\includegraphics[width=0.42\textwidth]{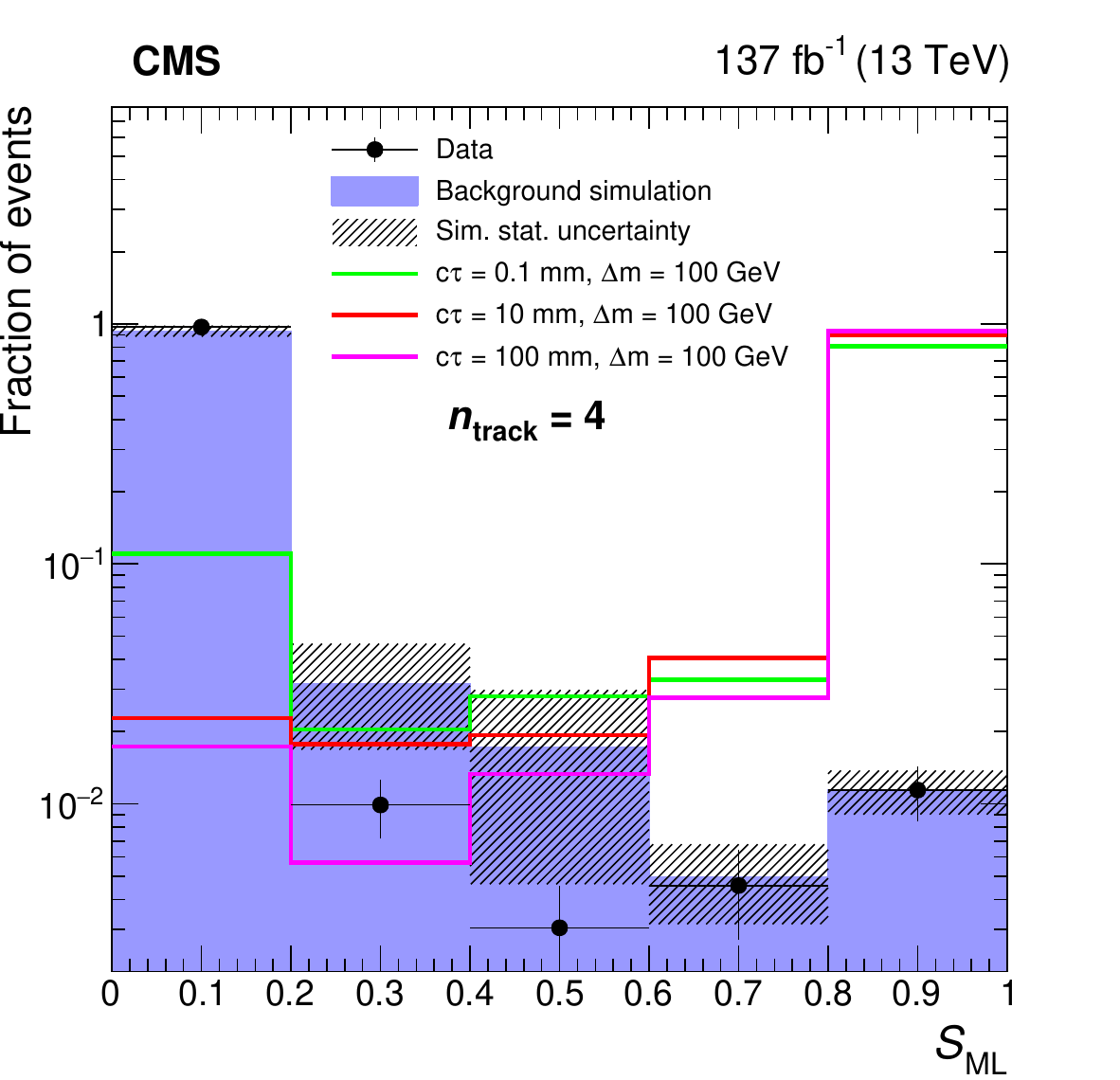}
\includegraphics[width=0.42\textwidth]{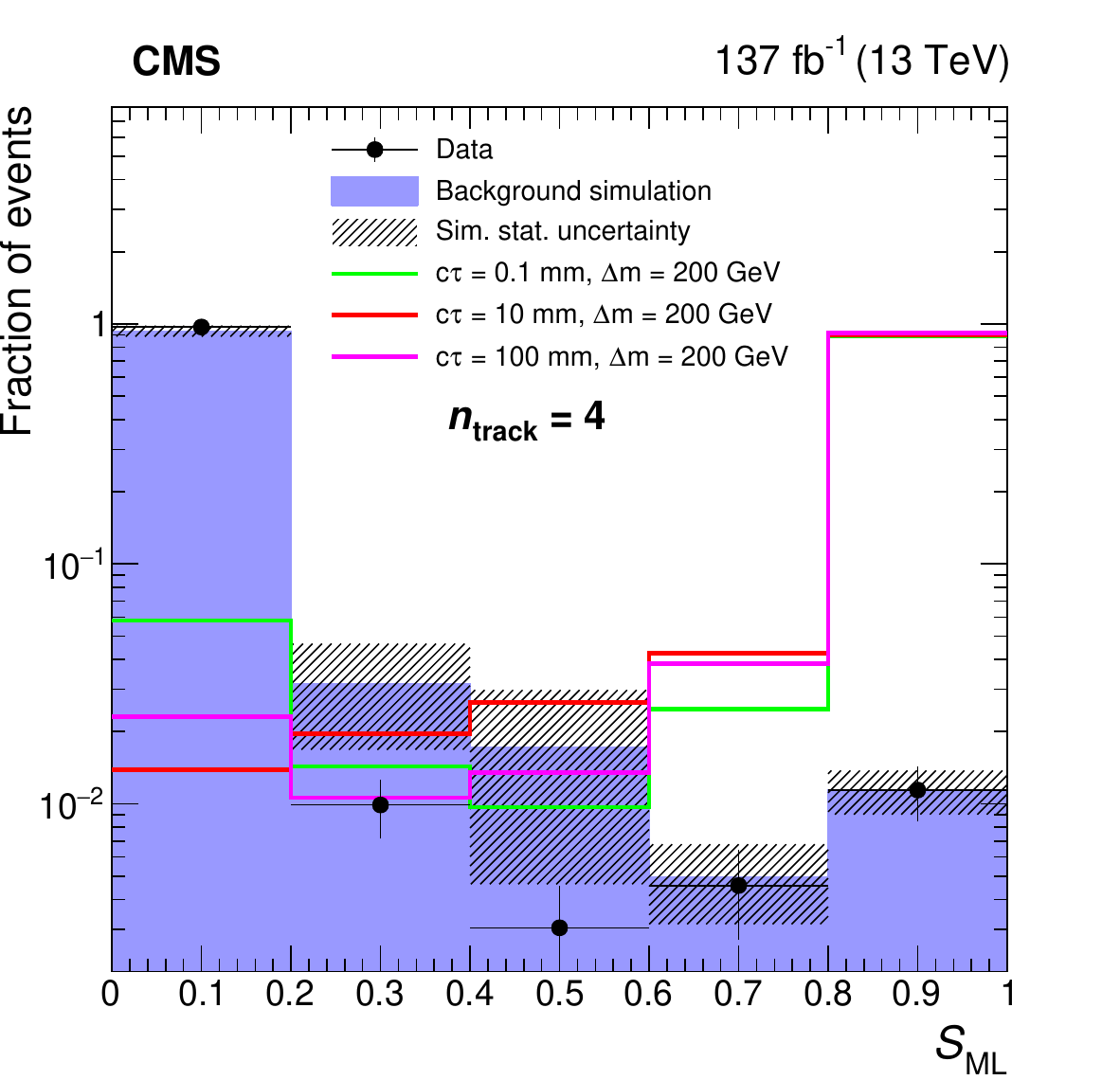}
\includegraphics[width=0.42\textwidth]{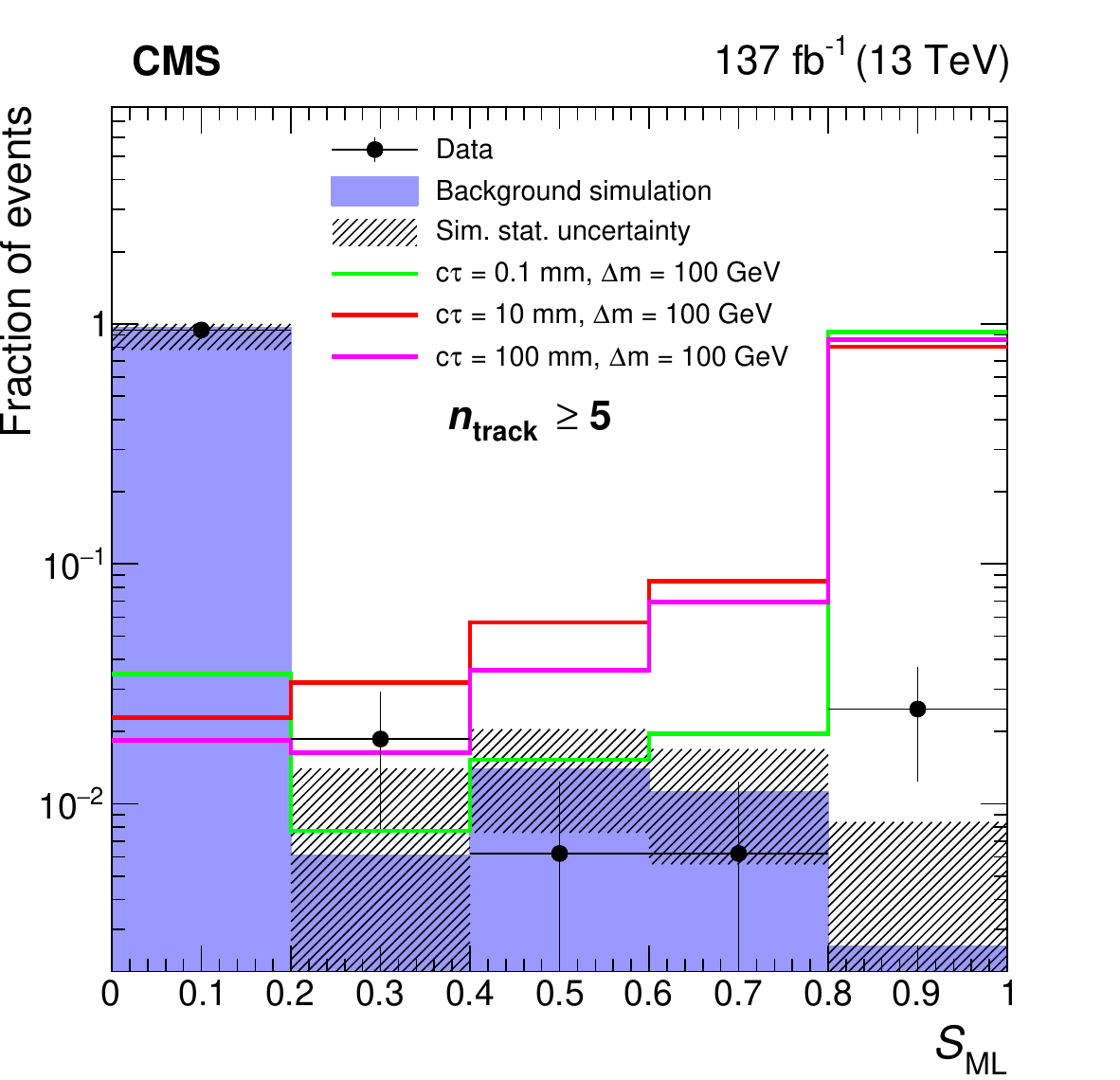}
\includegraphics[width=0.42\textwidth]{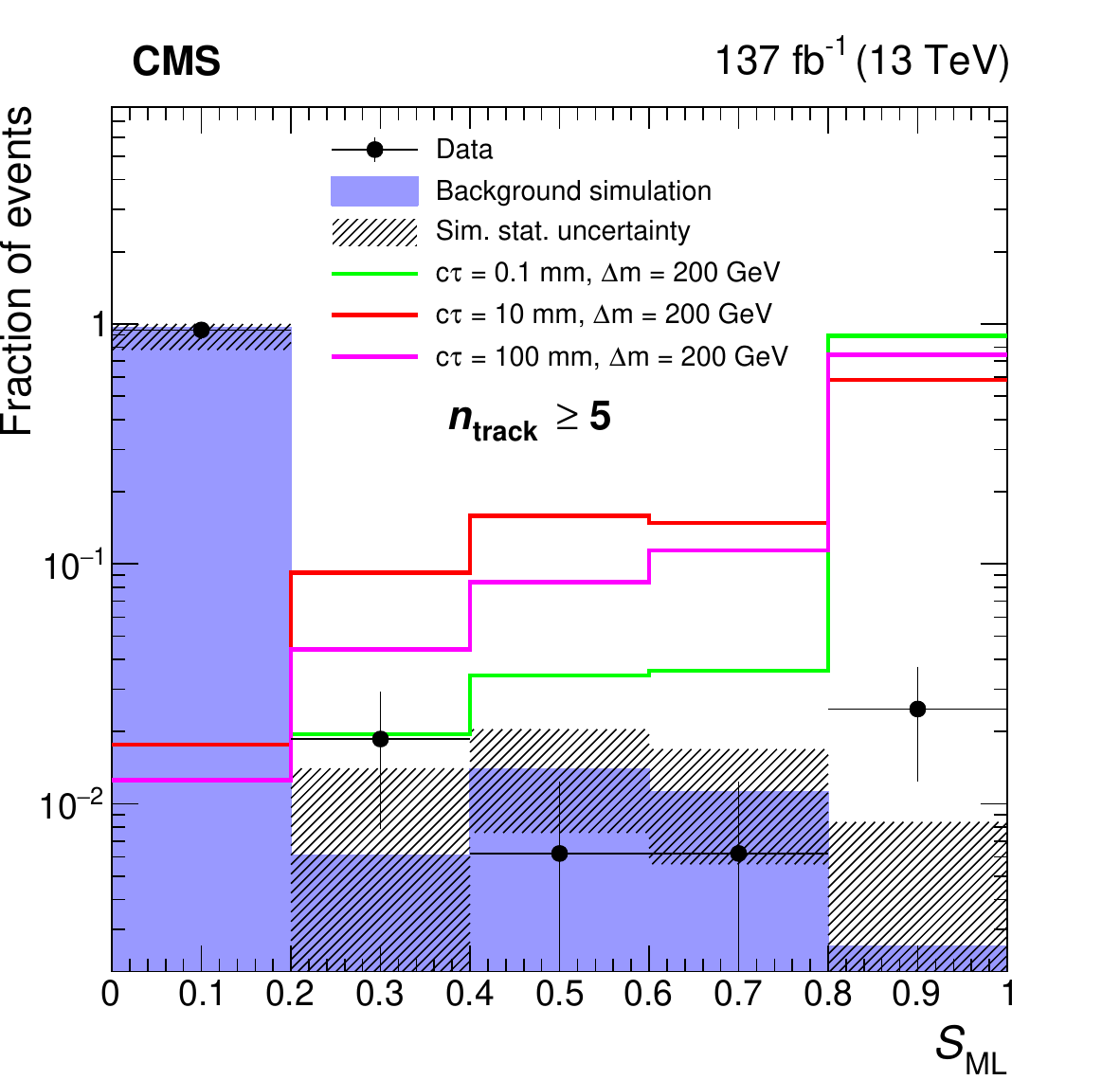}
\caption{Distributions of \mlscore for data, simulated background, and signal. Events with \ntk of 3 (upper), 4 (middle), and $\geq$5 (lower) are shown individually. The background simulation is shown for illustration and it is not used in the search.
The distributions are shown for split-SUSY signals with a gluino mass of 2000\GeV and neutralino mass of 1900\GeV (left) and 1800\GeV (right). Different gluino proper decay lengths and mass difference between the gluino and neutralino, are shown as $c\tau$ and $\Delta m$ in the legend. All distributions are normalized to unity.
}
\label{fig:mlscore_bkg_sig}
\end{figure*}

The testing events are used to evaluate the performance of the IN. The \mlscore distributions in simulated background events and split-SUSY signal events with $\ntk = 3$, 4, and $\geq$5 are shown in Fig.~\ref{fig:mlscore_bkg_sig}. The significant differences between the \mlscore distributions of background and signal samples show the strong discriminating power of the IN. Tagging events with $\mlscore > 0.2$ as signal-like events correctly identifies approximately 97\% of signal events while rejecting approximately 97\% of background events.

The \mlscore distributions of signal events with $\ntk \geq 5$ have an increased likelihood around \mlscore of 0.4 that is caused by the decorrelation term of the IN loss function. During the training, background events with larger \ntk are more likely to obtain higher \mlscore without the DisCo regulation. To achieve the decorrelation, the IN tends to assign a smaller \mlscore value to events with larger \ntk. Given that events from signal samples with larger mass splittings and lifetimes are more likely to have well-reconstructed vertices with larger \ntk, they are also more likely to obtain a smaller \mlscore.

\begin{figure}[hbtp]
\centering
\includegraphics[width=0.48\textwidth]{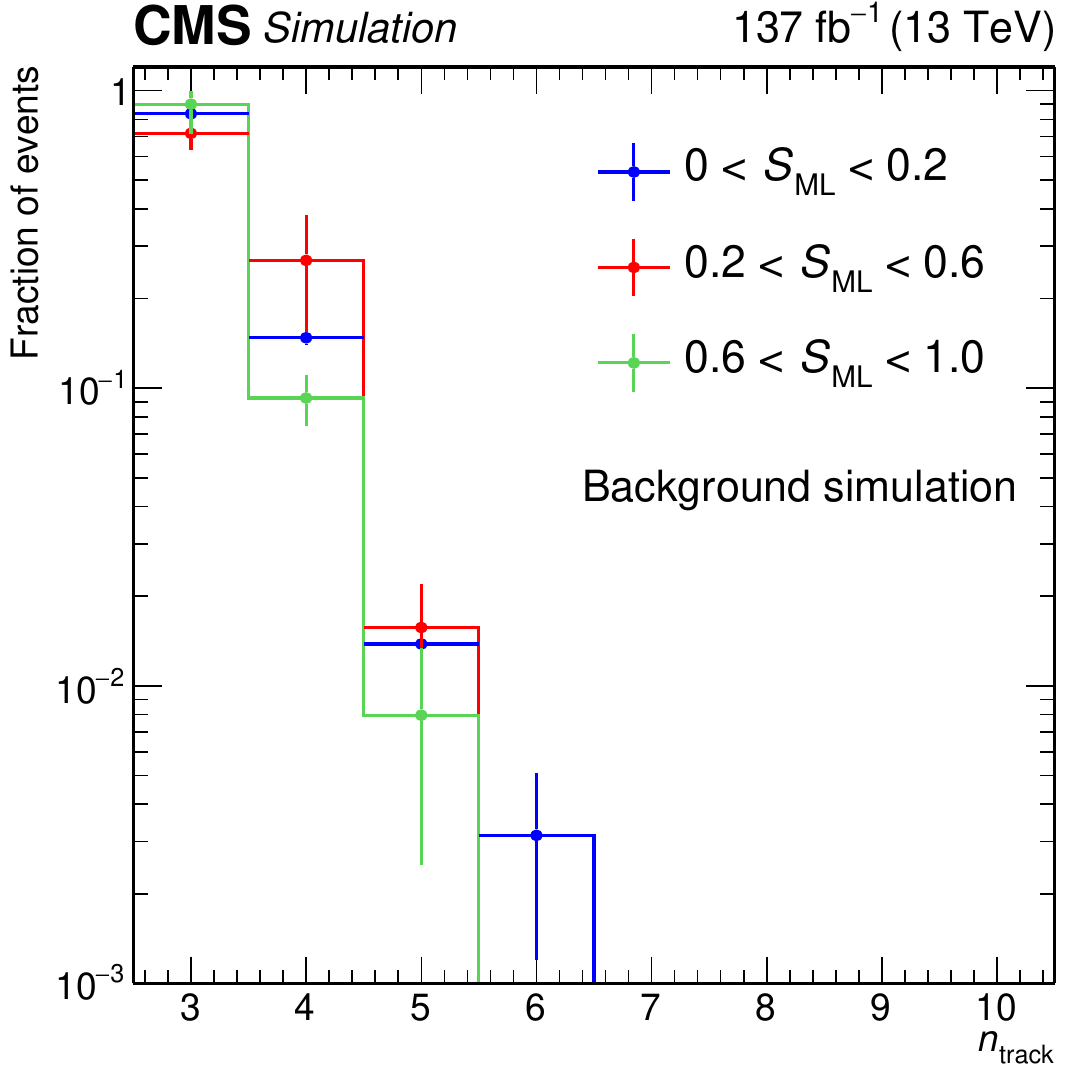}
\caption{The distribution of \ntk in different \mlscore regions for simulated background events. Events with $0 < \mlscore < 0.2$ (blue), $0.2 < \mlscore < 0.6$ (red), and $0.6 < \mlscore < 1.0$ (green) are compared. All distributions are normalized to unity. The similar \ntk distributions demonstrate that \ntk and \mlscore are decorrelated.
}
\label{fig:ntk_MLscore_regions}
\end{figure}

To estimate the effect of decorrelation between \mlscore and \ntk, we divide events into three regions depending on the \mlscore: $0 < \mlscore < 0.2$, $0.2 < \mlscore < 0.6$, and $0.6 < \mlscore < 1.0$. The \ntk distributions in different regions are compared, as shown in Fig.~\ref{fig:ntk_MLscore_regions}. As seen in the distributions, background events have similar \ntk distributions in different \mlscore regions, which demonstrates the success of the decorrelation technique. 

\section{Event selection and signal efficiency}
\label{sec:vtx_ml_eff}

The signal region contains all events that pass the event preselection, have at least one displaced vertex with $\ntk \geq 5$, and have $\mlscore > 0.2$. The overall signal selection efficiency, which is defined as the fraction of generated signal events that pass the signal region selection, is directly related to the track reconstruction efficiency, vertex reconstruction efficiency, and ML tagging efficiency. It is important to understand these efficiencies in data and simulation and account for any observed differences. The procedures are based on methods similar to those of Ref.~\cite{EXO-19-013}. The simulated background processes described in Section~\ref{sec:datasets} are used in the studies. 

\begin{figure}[bt]
\centering
\includegraphics[width=0.48\textwidth]{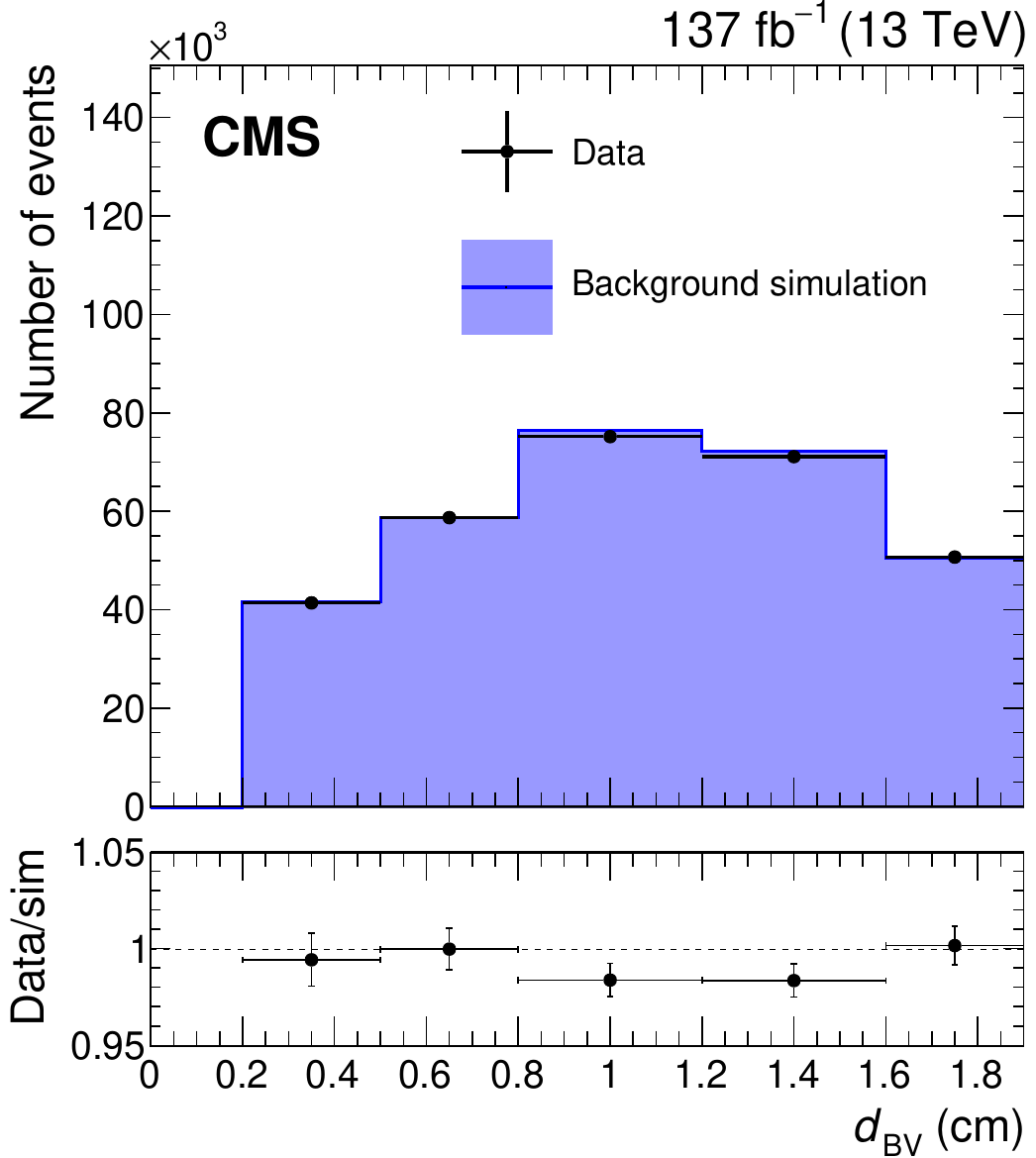}
\caption{The distribution of \dbv in \PKzS vertices in data (black) and simulation (purple). The lower panel shows the ratio between data and simulation.
}
\label{fig:k0dbv_2017}
\end{figure}

The track reconstruction efficiency is studied as a function of \dbv in data and simulation using events enriched in $\PKzS\to\PGpm\PGpp$ decays. The \PKzS decays are selected by requiring events that satisfy the preselection and contain a pair of tracks that form a well-reconstructed and displaced vertex. The vertex is required to be located inside the beam pipe, and the invariant mass of the two tracks associated with the vertex is required to be between 0.490 and 0.505\GeV when assuming the charged-\PGp mass hypothesis.
The \dbv distribution of resulting \PKzS vertices is compared between data and simulation to determine potential corrections to the signal simulation. 
The \dbv distribution in simulation is normalized such that the number of \PKzS vertices is consistent with data in the 0.5--0.8\mm region, where tracking efficiency is known to be well modeled and the number of \PKzS vertices results in a sufficiently small statistical uncertainty in the normalization. 
The number of reconstructed \PKzS vertices in data and simulation as a function of \dbv is shown in Fig.~\ref{fig:k0dbv_2017}.
Given that each \PKzS vertex is composed of two tracks and failing to reconstruct one track is the dominant cause of the failure to reconstruct a \PKzS vertex, the track reconstruction efficiency is proportional to the square root of the \PKzS vertex reconstruction efficiency. 
The ratio of data-to-simulation track reconstruction efficiency is found to be between 98 and 100\% in all data-taking periods and bins of \dbv. These ratios are used to derive systematic uncertainties in the signal efficiency using the procedure described in Section~\ref{sec:sig_syst}.

The vertex reconstruction and ML tagging efficiencies are studied by artificially displacing tracks in data and simulated background events to produce displaced vertices that mimic displaced vertices generated by LLP decays in signal events. Events are required to satisfy the preselection, which results in a sample dominated by background processes because of the lack of track displacement and vertex reconstruction requirements. All simulated background processes described in Section~\ref{sec:datasets} are used in the study. From the selected events, jets with $\pt>20\GeV$ and at least three matched PF candidate tracks are considered for further study. These jets are categorized as \PQb or light jets based on whether they pass or fail the ``tight'' working point of the \textsc{DeepJet} algorithm~\cite{Bols:2020bkb}, which has a \PQb tagging efficiency of 55--65\% in QCD events with $\pt>30\GeV$ and a misidentification rate for light jets of 0.1\%. 
To mimic the topology of signal events, two light jets are randomly selected and tracks associated with the selected jets are displaced simultaneously from the primary vertex. The direction of the displacement is determined by the vector sum of the selected jet momenta, which is then subjected to a Gaussian smear with a standard deviation of 2.0\unit{rad} on both $\phi$ and the polar angle $\theta$, while the magnitude of the displacement is randomly sampled from an exponential distribution with $c\tau$ of 10\mm, which reproduces the gluino decays in the simulated signal events. Corrections to the angular separation and the number of artificially displaced tracks per jet, derived from the ratios of those distributions between signal and events with artificial track displacements, are applied to ensure they describe signal vertices well. 

\begin{figure*}[hbtp]
\centering
\includegraphics[width=0.49\textwidth]{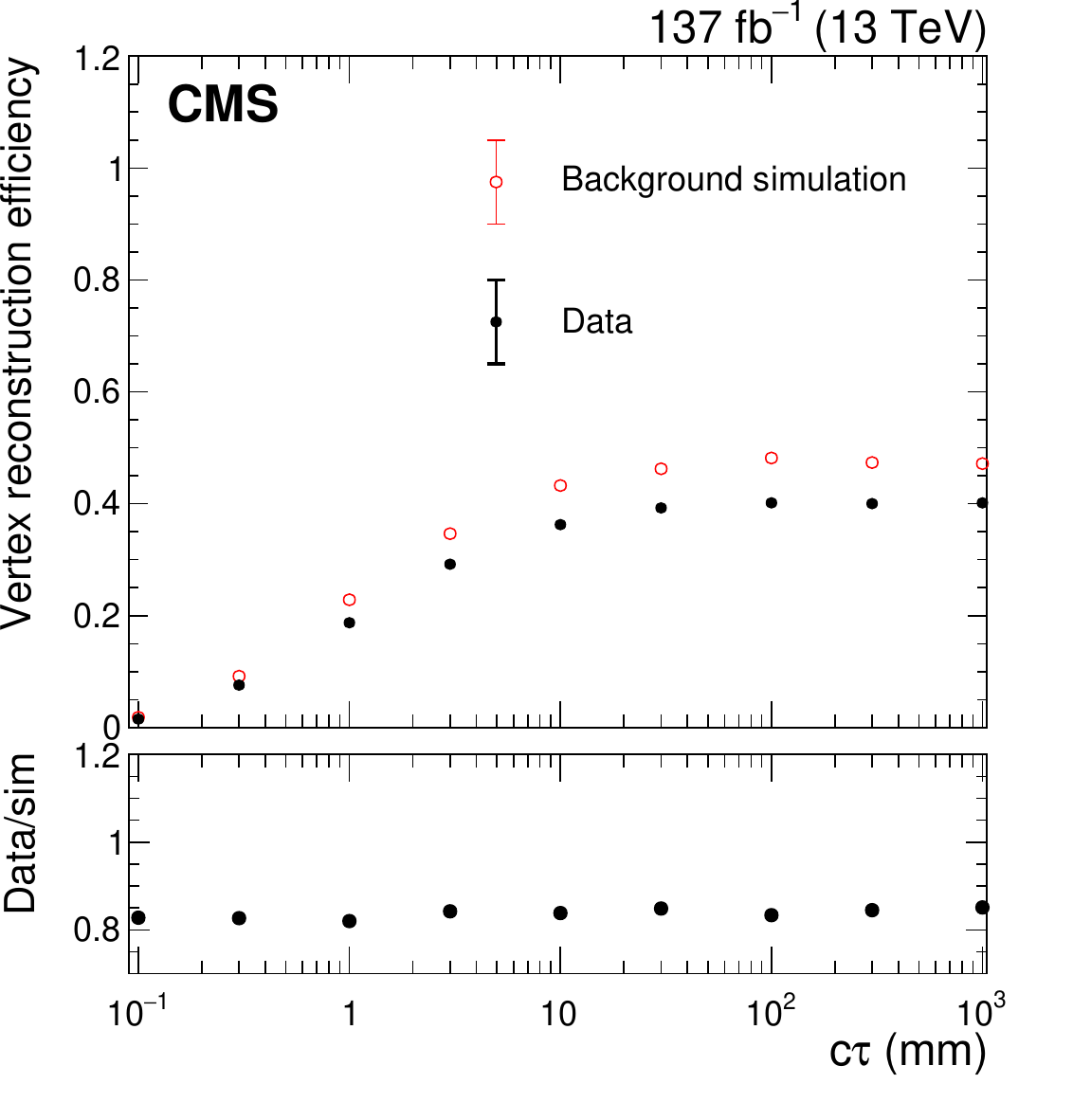}
\includegraphics[width=0.49\textwidth]{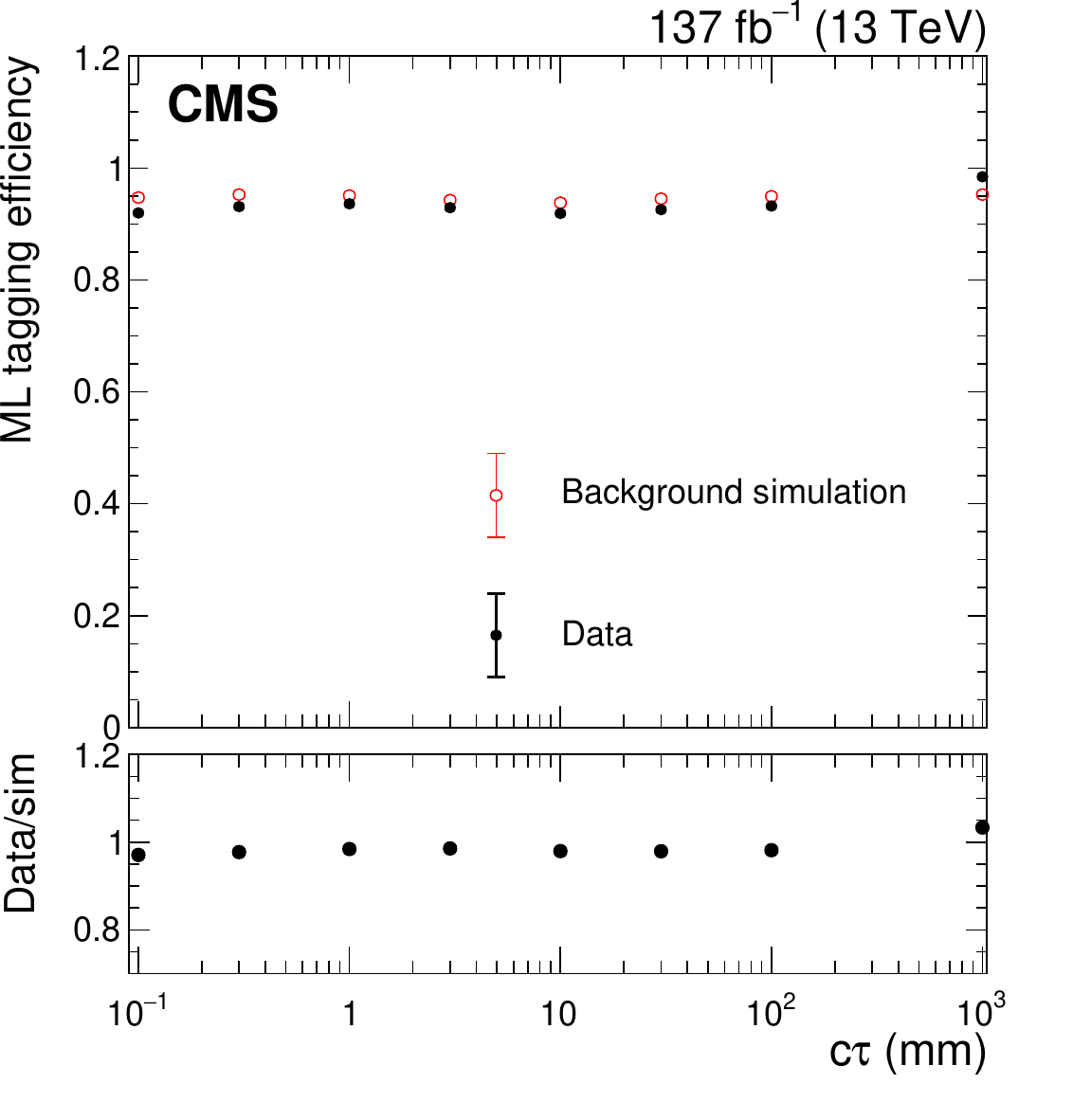}
\caption{The vertex reconstruction efficiency (left) and ML tagging efficiency (right) for artificially displaced vertices in data (black) and simulation (red). In this example, the artificially displaced vertices are corrected to mimic split-SUSY signal events with gluino mass of 2000\GeV and neutralino mass of 1800\GeV. The uncertainties are too small to be visible in the plot.
}
\label{fig:trackmover_2017}
\end{figure*}

The vertex reconstruction and ML algorithms are then applied to the events with artificially displaced tracks. The vertex reconstruction efficiency is calculated as the fraction of all artificially displaced vertices that are reconstructed within 200\mum of the expected position and satisfy the vertex selections described in Section~\ref{sec:vtxreco}. The ML tagging efficiency is calculated as the fraction of events with a reconstructed vertex that have \mlscore larger than 0.2. 
The efficiencies are measured for both data and simulated events, and are compared in Fig.~\ref{fig:trackmover_2017}. 
The vertex reconstruction efficiency ratios between data and simulation range from 80 to 97\%, depending on different mass splittings, while the ML tagging efficiency ratios range from 91 to 100\% for different mass splittings. In contrast, the data-to-simulation efficiency ratios differ by less than a few percent between samples with different gluino masses and $c\tau$ values. Therefore, the efficiency ratio value used to calculate correction factors and systematic uncertainties is only dependent on the mass splitting and data-taking period.

The simulated signal event yield is corrected by scaling the vertex reconstruction efficiency ratio to reproduce the expected behavior in data. The differences between unity and the vertex efficiency and ML tagging efficiency ratios are used as systematic uncertainties following the procedures described in Section~\ref{sec:syst}.

\section{Background estimation}
\label{sec:bkgest}

\begin{figure}[hbtp]
\centering
\includegraphics[width=0.49\textwidth]{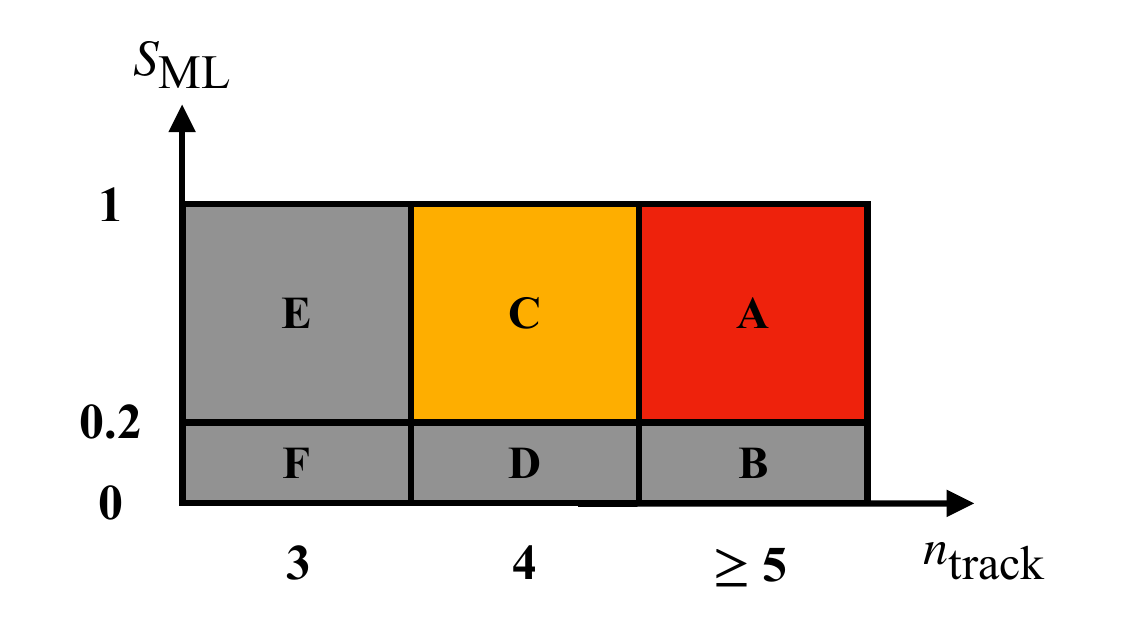}
\caption{A schematic diagram of the signal (red), validation (yellow), and control (gray) regions. The letter in each box corresponds to the region label described in the text.
}
\label{fig:ABCD_cartoon}
\end{figure}

The background events that pass the signal region selection are expected to be dominated by events in which displaced vertices are formed by random crossings of otherwise unrelated tracks. We estimate the number of such events and validate the estimation procedure using regions in data that are chosen to be orthogonal to the signal region and dominated by similar events. The control and validation regions are constructed from all events that pass the event preselection and have at least one reconstructed displaced vertex with $\ntk \geq 3$ but do not pass the signal region selections ($\ntk \geq 5$ and $\mlscore > 0.2$). 
Compared with background events in the signal region, background vertices with lower numbers of tracks are similar because they result from the same effect, \ie, unrelated tracks randomly crossing each other. However, the possibility to form a high track multiplicity background vertex is lower because it is less likely to have more tracks crossing the same spatial point by coincidence. 
The events passing these selections are then separated into three bins of \ntk ($\ntk = 3$, 4, or $\geq$5), and each \ntk bin is further divided into two bins depending on whether $\mlscore > 0.2$ is satisfied. Figure~\ref{fig:ABCD_cartoon} illustrates how the control regions (labeled B, D, E, and F), validation region (C), and signal region (A) are arranged in the plane defined by \ntk and \mlscore. All control and validation regions are dominated by background events from \ttbar and QCD processes.

The background estimation method takes advantage of the statistical independence of \mlscore and \ntk, which is ensured by the DisCo method. Because the fraction of background events that satisfy \mlscore is independent of the value of \ntk, the ratio of events in region E to events in region F should be consistent with the equivalent ratios of regions C to D and regions A to B, within the statistical uncertainty. The number of background events in the validation and signal regions can therefore be estimated as
\begin{equation}
\label{eq:sr}
N_{\text{bkg}}^{\text{A}}=\frac{N_{\text{bkg}}^{\text{B}} N_{\text{bkg}}^{\text{E}}}{N_{\text{bkg}}^{\text{F}}},
\end{equation}
\begin{equation}
\label{eq:vr}
N_{\text{bkg}}^{\text{C}}=\frac{N_{\text{bkg}}^{\text{D}} N_{\text{bkg}}^{\text{E}}}{N_{\text{bkg}}^{\text{F}}}.
\end{equation}
To validate the background estimation procedure that is used to estimate the number of events in region A using Eq.~(\ref{eq:sr}), we predict the number of events in region C using Eq.~(\ref{eq:vr}) and compare this prediction to the observed number of events in region C. The prediction and observation agree within statistical uncertainties.

\section{Systematic uncertainties}
\label{sec:syst}

\subsection{Systematic uncertainties affecting signal}
\label{sec:sig_syst}
Systematic uncertainties associated with signal reconstruction are due to uncertainties in the modeling of the track reconstruction efficiency, vertex reconstruction efficiency, ML tagging efficiency, \ptmiss, choice of PDF, trigger efficiency, pileup, integrated luminosity, and running conditions. The dominant source of systematic uncertainty is associated with the choice of PDF, followed by the modeling of vertex reconstruction efficiency.

The track reconstruction efficiency impacts the signal efficiency because $\ntk \geq 5$ is one of the signal region selections. In cases where an LLP decays to exactly five charged particles, failure to reconstruct even one of the five tracks will cause the vertex to fall outside of the signal region, and the impact of track reconstruction on such vertices is calculated as the fifth power of the track reconstruction efficiency. On the other hand, for vertices with more than five tracks, the failure to reconstruct any of the tracks will not necessarily cause the vertex to be excluded from the signal region, as long as there are at least five tracks remaining. Since the track reconstruction efficiency in data and simulation agree within 2\%, the impact of track reconstruction efficiency is negligible for those vertices. We apply a systematic uncertainty equal to the differences in the fifth power of the track reconstruction efficiency observed in the study described in Section~\ref{sec:vtx_ml_eff}. The assigned uncertainties range from 6 to 21\%, depending on signal mass, signal lifetime, and data-taking period.

In addition, a systematic uncertainty associated with the correction to the vertex reconstruction efficiency, described in Section~\ref{sec:vtx_ml_eff}, is assigned to account for the potential remaining differences between data and simulation, discrepancies between artificially displaced events and simulated signal events, differences between different gluino masses and proper decay lengths, and other small effects. The systematic uncertainties are taken to be the differences in vertex reconstruction efficiency between data and simulation observed in the study described in Section~\ref{sec:vtx_ml_eff}, with values ranging from 3 to 20\%.

Similarly, a systematic uncertainty associated with ML tagging efficiency is assigned to account for any differences in the behavior of the ML algorithm between data and simulation. The magnitude of this uncertainty obtained as the difference of ML tagging efficiency between data and simulation in the study described in Section~\ref{sec:vtx_ml_eff}, ranges from negligibly small to 24\%.

Potential mismodeling of jet and unclustered energy scales and jet energy corrections impact the \ptmiss value, and therefore affect the \ptmiss selection efficiency. To account for this effect, we assign a systematic uncertainty of up to 8\%, derived from the change in signal efficiency when varying the parameters associated with the mismodeling by 1 standard deviation about their mean.
The variations of the PDF could result in changes in the selection efficiency in the signal region. The associated uncertainty is evaluated by reweighting simulated signal events using the variation observed between NNPDF replica sets~\cite{Butterworth:2015oua}. For most of the signal samples, the resulting PDF uncertainty is within 20\%. However, for signal samples with very low signal efficiency, as the ones with mass splittings below 50\GeV and $c\tau$ values of 0.1 or 1000\mm, the PDF uncertainty is affected by statistical fluctuations and reaches values as large as 85\%.
The trigger efficiency systematic uncertainty is determined by its statistical uncertainty and the variation when using different data sets, resulting in an uncertainty of 1--6\%. The uncertainty associated with the modeling of the pileup distribution is 2--15\%. The integrated luminosity measurements for the 2016, 2017, and 2018 data-taking years have 1--3\% individual uncertainties~\cite{CMS-LUM-17-003,CMS-PAS-LUM-17-004,CMS-PAS-LUM-18-002}, while the overall uncertainty for the 2016--2018 period is 1.6\%. 
During 2016 and 2017 data-taking periods, a gradual timing shift in the ECAL was not correctly accounted for in the L1 trigger, resulting in an efficiency drop for events with significant ECAL energy distributed in the high $\eta$ region. This effect results in an uncertainty in the signal yield up to 1\%.

A summary of systematic uncertainties that affect the signal yield is presented in Table~\ref{tab:syst_sig}. The overall systematic uncertainty is calculated by summing individual components in quadrature under the assumption that there are no correlations between them. 

\begin{table}[htbp!]
    \centering
    \topcaption{Summary of systematic uncertainties that affect the signal yield. The magnitude of each systematic varies by data-taking period and signal parameters, so a range of values is given in each case.
    }
    \begin{scotch}{l c}
    Systematic uncertainty  & Magnitude (\%)   \\
    \hline
    Track reconstruction    & 6--21  \\
    Vertex reconstruction   & 3--20 \\
    ML tagging              & $\leq$24   \\
    \ptvecmiss selection    & $\leq$8 \\
    PDF uncertainty         & 1--85 \\
    Trigger efficiency      & 1--6    \\
    Pileup                  & 2--15   \\
    Integrated luminosity   & 1--3   \\
    L1 trigger inefficiency & $\leq$1   \\[\cmsTabSkip]
    Total                 & 8--91 
    \end{scotch}
    \label{tab:syst_sig}
\end{table}

\subsection{Systematic uncertainties in background estimation}
Since the background is estimated from data, potential mismodeling of the simulation does not have an impact. The number of events in the signal region is predicted using the number of events in control regions B, E, and F, which are based on 3-track and 5-track control regions. Given that the two variables that define the search regions are statistically independent, the prediction of the number of background events in the signal region based on Eq.~(\ref{eq:sr}) should be the same using 3-track control regions (E and F) or 4-track control regions (C and D). To validate this assumption, we compare two predictions of the number of background events in the signal region, one based on 3-track control regions and the other based on 4-track control regions. We find that the two predictions are compatible within the statistical uncertainty and therefore do not assign an additional systematic uncertainty in the background estimation.

\section{Results and statistical interpretation}
\label{sec:results}

A maximum likelihood fit under the background-only hypothesis is performed on all search regions. Systematic uncertainties described in Section~\ref{sec:syst} are taken to be uncorrelated and used as nuisance parameters in the fit. The predicted and observed numbers of events in all search regions after the fit are reported in Table~\ref{tab:event_yield_dataRunII}. In the signal region, nine events are observed while $5.2 \pm 0.5$ events are predicted, which corresponds to a $p$-value of 0.089. No significant discrepancy between the background-only fit prediction and the observation is seen. 

\begin{table}[htbp!]
    \centering
    \topcaption{
    Number of predicted and observed events in the control, validation, and search regions. Predictions are calculated using Eqs.~(\ref{eq:sr}) and (\ref{eq:vr}) and fitting the data under the background-only hypothesis. Regions are organized by \mlscore and \ntk values, and region names corresponding to Fig.~\ref{fig:ABCD_cartoon} are given in parentheses. The predicted number of events that pass the \mlscore selection and the observed number of events that pass or fail the \mlscore selection are shown in separate rows.}
    \begin{scotch}{c r r r}
                               & $\ntk = 3$ & $\ntk = 4$         & $\ntk \geq 5$ \\ 
    \hline
    Predicted $\mlscore > 0.2$ & \NA~(E)    & $38.0 \pm 6.0$ (C) & $5.2 \pm 0.5$ (A) \\ 
    Observed $\mlscore > 0.2$  & 203 (E)    & 38 (C)             & 9 (A) \\ 
    Observed $\mlscore < 0.2$  & 6327 (F)   & 1276 (D)           & 152 (B)
    \end{scotch}
    \label{tab:event_yield_dataRunII}
\end{table}

Based on the observed number of events and the background prediction, constraints are set on parameters of the benchmark signal samples used in the search. Specifically, the 95\% confidence level (\CL) upper limit on the product of the production cross section and the square of the branching fraction ($\sigma \mathcal{B}^{2}$) of the decay mode is estimated using the \CLs method~\cite{CLS1,CLS2,CMS:2024onh}. The signal yields are determined from simulated signal events, with all the corrections applied; the number of background events is determined purely from data. Data sets from different data-taking
periods are treated separately and combined during the final fit. The upper limits are compared with the theoretical prediction of the production cross section and uncertainty~\cite{Borschensky:2014cia} calculated at NNLO+NNLL precision~\cite{Beenakker:2016lwe} to exclude regions of the parameter space of the benchmark signal models. The branching fraction for the gluino decay to the targeted final state in split SUSY and GMSB SUSY is assumed to be 100\% for all signal samples.

\begin{figure}[hbtp]
\centering
\includegraphics[width=0.49\textwidth]{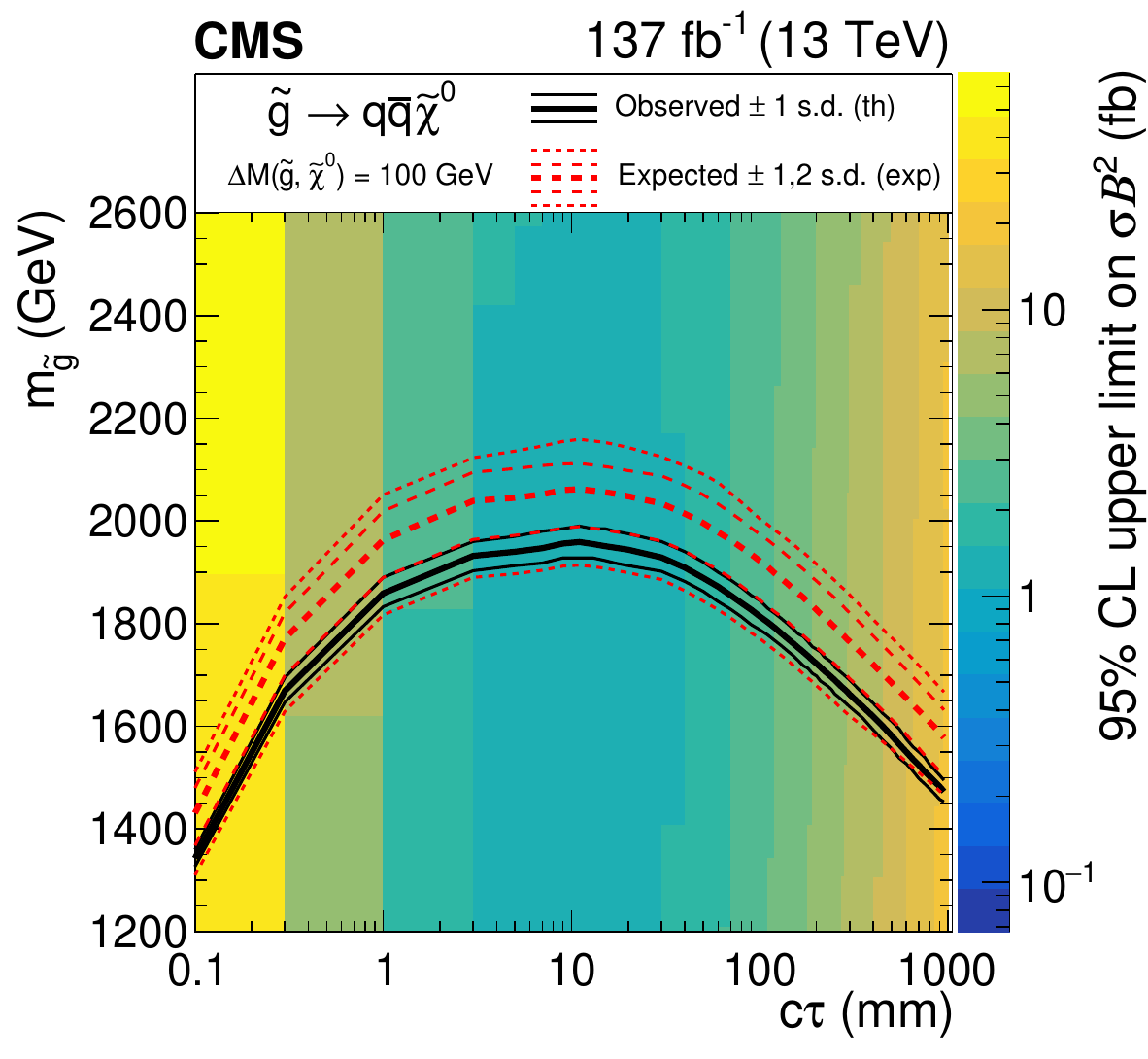}
\includegraphics[width=0.49\textwidth]{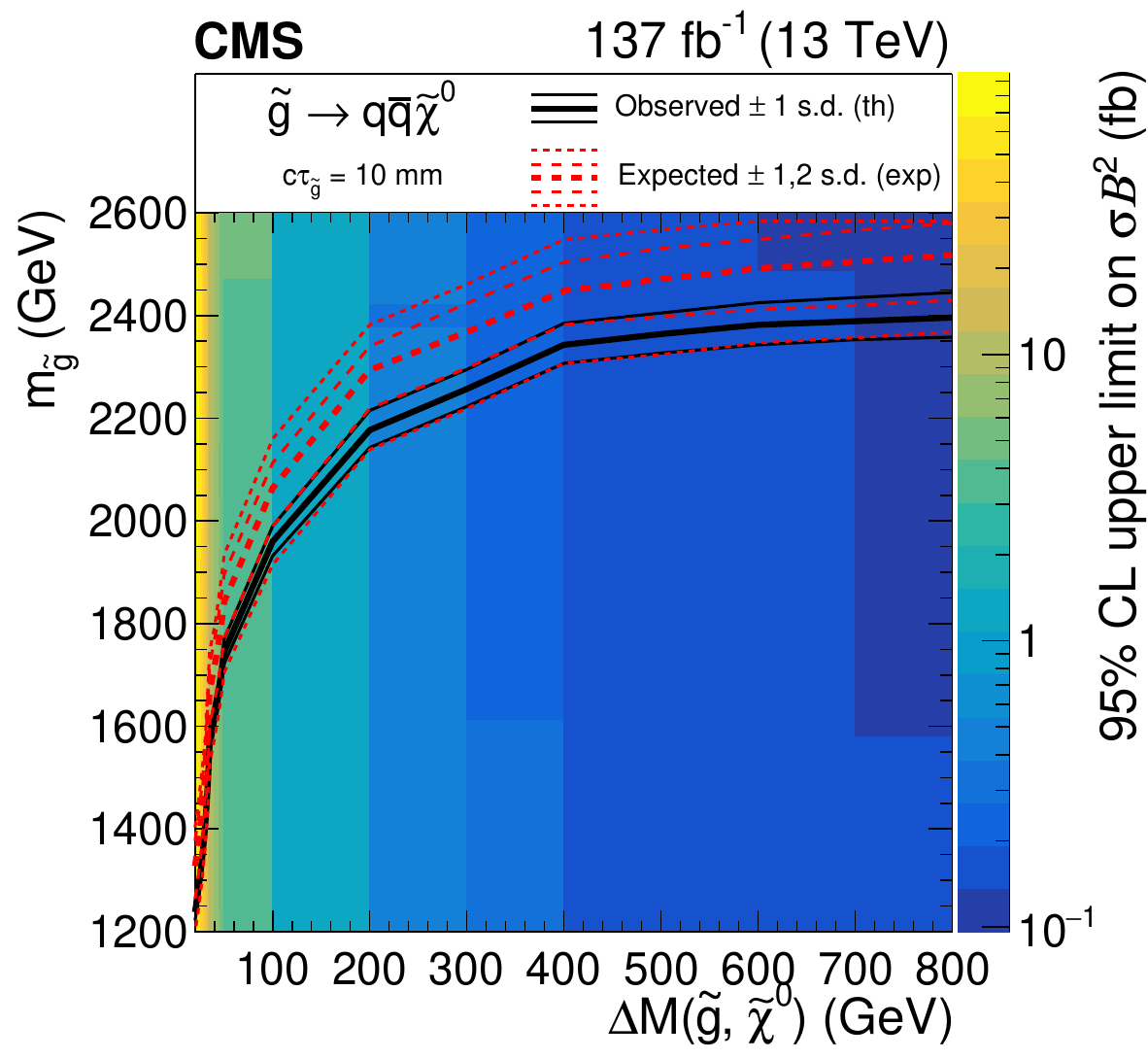}
\caption{\cmsLLeft: the 95\% \CLs upper limit on the product of the cross section and branching fraction squared for the split-SUSY signal model with a mass splitting of 100\GeV, shown as a function of gluino mass and $c\tau$. \cmsRRight: the 95\% \CLs upper limit on the product of the cross section and branching fraction squared for the split-SUSY model with a $c\tau$ of 10\mm, shown as a function of gluino mass and mass splitting.
For both plots, the observed (solid black) and expected (dashed red) exclusion curves are shown.
}
\label{fig:limits2d_splitSUSY_RunII}
\end{figure}

\begin{figure}[hbtp]
\centering
\includegraphics[width=0.49\textwidth]{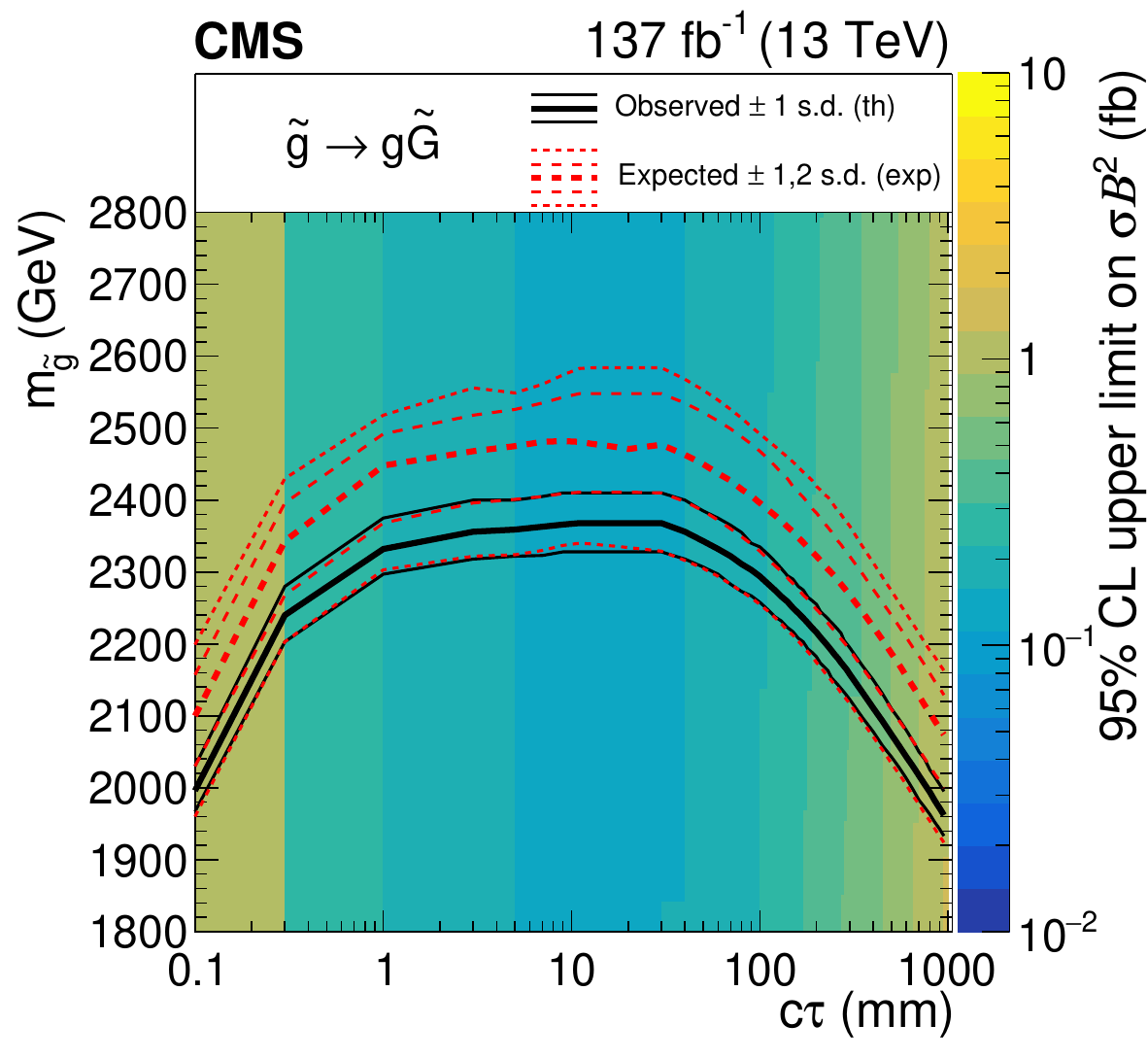}
\caption{The 95\% \CLs upper limit on the product of the cross section and branching fraction squared for the GMSB SUSY signal model, shown as a function of gluino mass and $c\tau$.
The observed (solid black) and expected (dashed red) exclusion curves are shown.
}
\label{fig:limits2d_gluinoGMSB_RunII}
\end{figure}

Figure~\ref{fig:limits2d_splitSUSY_RunII} shows the upper limits and exclusion curves for the split-SUSY model as a function of gluino mass, gluino $c\tau$, and gluino-neutralino mass splitting. The search is most sensitive when gluino $c\tau$ is approximately 10\mm and the gluino-neutralino mass splitting is large, but the sensitivity extends to gluino $c\tau$ values from 0.1 to 1000\mm and mass splittings as low as 20\GeV. The decreasing sensitivity at the low $c\tau$ values is driven by the decreasing displaced vertex reconstruction efficiency, while the decreasing sensitivity at highest $c\tau$ values is driven by the constraint that the displaced vertex must be within the CMS beam pipe. Figure~\ref{fig:limits2d_gluinoGMSB_RunII} shows the upper limits and exclusion curves for the GMSB SUSY model as a function of gluino mass and $c\tau$, which exhibit a similar dependence on gluino lifetime to those associated with the split-SUSY model.

The observed upper limits are weaker than the expected by about 1.35 standard deviations since more events are observed in the signal region compared to the expectation. As this search uses only one signal region, the upper limits for different signal parameters are calculated using the same region, which makes the difference between observed and expected upper limits similar for different signal parameters. 
For the split-SUSY benchmark signal model, the search excludes gluinos with $c\tau$ in the range 1--100\mm and masses below 1800\GeV for mass splittings of 100\GeV with respect to the neutralino. Also, for mass splittings above 50\GeV, the search excludes gluinos with masses below 1600\GeV and $c\tau$ between 1--30\mm. For gluino masses of 1400\GeV and neutralino masses of 1300\GeV, this search excludes production cross sections that are a factor of 2--100 times smaller than those excluded by Ref.~\cite{ATLAS:2017tny} and 10--100 times smaller than those excluded by Ref.~\cite{CMS:2018ncl}, depending on the gluino lifetime. 
For the GMSB SUSY signal model, the search achieves upper limits on the product of the cross section and branching fraction squared as low as 1\unit{fb} for gluinos with $c\tau$ ranging from 0.1 to 1000\mm and excludes gluinos with $c\tau$ in the range 0.3--100\mm and masses below 2240\GeV.

\section{Summary}
\label{sec:summary}

A search for the production of long-lived particles that decay to at least one displaced vertex with missing transverse momentum in proton-proton collisions at a center-of-mass energy of 13\TeV collected by the CMS detector has been presented. The analysis extends the previous CMS search~\cite{EXO-19-013} by improving the sensitivity to events with low total jet energy, targeting events with as few as one displaced vertex, and introducing a dedicated machine learning algorithm that reduces the number of background events in the signal region by 94\%. Split supersymmetry (SUSY) and gauge-mediated SUSY breaking are used as benchmark signal models for statistical interpretations in this search. 

At 95\% confidence level, the search excludes long-lived gluinos predicted by the split-SUSY model with masses below 1800\GeV and mean proper decay lengths in the range of 1 to 100\mm, when the mass splitting is 100\GeV. For mass splittings above 50\GeV, gluinos with masses below 1600\GeV and mean proper decay lengths between 1 and 30\mm are excluded.
For the gauge-mediated SUSY breaking model, gluinos with masses below 2200\GeV and mean proper decay lengths between 0.3 and 100\mm are excluded. 
This search is the first CMS search that shows sensitivity to hadronically decaying long-lived particles from signals with mass differences between the gluino and neutralino below 100\GeV. It sets the most stringent limits to date for split-SUSY models and for GMSB gluinos with proper decay length less than 6\mm.

\begin{acknowledgments}
We congratulate our colleagues in the CERN accelerator departments for the excellent performance of the LHC and thank the technical and administrative staffs at CERN and at other CMS institutes for their contributions to the success of the CMS effort. In addition, we gratefully acknowledge the computing centers and personnel of the Worldwide LHC Computing Grid and other centers for delivering so effectively the computing infrastructure essential to our analyses. Finally, we acknowledge the enduring support for the construction and operation of the LHC, the CMS detector, and the supporting computing infrastructure provided by the following funding agencies: SC (Armenia), BMBWF and FWF (Austria); FNRS and FWO (Belgium); CNPq, CAPES, FAPERJ, FAPERGS, and FAPESP (Brazil); MES and BNSF (Bulgaria); CERN; CAS, MoST, and NSFC (China); MINCIENCIAS (Colombia); MSES and CSF (Croatia); RIF (Cyprus); SENESCYT (Ecuador); ERC PRG, RVTT3 and MoER TK202 (Estonia); Academy of Finland, MEC, and HIP (Finland); CEA and CNRS/IN2P3 (France); SRNSF (Georgia); BMBF, DFG, and HGF (Germany); GSRI (Greece); NKFIH (Hungary); DAE and DST (India); IPM (Iran); SFI (Ireland); INFN (Italy); MSIP and NRF (Republic of Korea); MES (Latvia); LMTLT (Lithuania); MOE and UM (Malaysia); BUAP, CINVESTAV, CONACYT, LNS, SEP, and UASLP-FAI (Mexico); MOS (Montenegro); MBIE (New Zealand); PAEC (Pakistan); MES and NSC (Poland); FCT (Portugal); MESTD (Serbia); MCIN/AEI and PCTI (Spain); MOSTR (Sri Lanka); Swiss Funding Agencies (Switzerland); MST (Taipei); MHESI and NSTDA (Thailand); TUBITAK and TENMAK (Turkey); NASU (Ukraine); STFC (United Kingdom); DOE and NSF (USA).
       
\hyphenation{Rachada-pisek} Individuals have received support from the Marie-Curie program and the European Research Council and Horizon 2020 Grant, contract Nos.\ 675440, 724704, 752730, 758316, 765710, 824093, 101115353, and COST Action CA16108 (European Union); the Leventis Foundation; the Alfred P.\ Sloan Foundation; the Alexander von Humboldt Foundation; the Science Committee, project no. 22rl-037 (Armenia); the Belgian Federal Science Policy Office; the Fonds pour la Formation \`a la Recherche dans l'Industrie et dans l'Agriculture (FRIA-Belgium); the Agentschap voor Innovatie door Wetenschap en Technologie (IWT-Belgium); the F.R.S.-FNRS and FWO (Belgium) under the ``Excellence of Science -- EOS" -- be.h project n.\ 30820817; the Beijing Municipal Science \& Technology Commission, No. Z191100007219010 and Fundamental Research Funds for the Central Universities (China); the Ministry of Education, Youth and Sports (MEYS) of the Czech Republic; the Shota Rustaveli National Science Foundation, grant FR-22-985 (Georgia); the Deutsche Forschungsgemeinschaft (DFG), under Germany's Excellence Strategy -- EXC 2121 ``Quantum Universe" -- 390833306, and under project number 400140256 - GRK2497; the Hellenic Foundation for Research and Innovation (HFRI), Project Number 2288 (Greece); the Hungarian Academy of Sciences, the New National Excellence Program - \'UNKP, the NKFIH research grants K 124845, K 124850, K 128713, K 128786, K 129058, K 131991, K 133046, K 138136, K 143460, K 143477, 2020-2.2.1-ED-2021-00181, and TKP2021-NKTA-64 (Hungary); the Council of Science and Industrial Research, India; ICSC -- National Research Center for High Performance Computing, Big Data and Quantum Computing, funded by the EU NexGeneration program (Italy); the Latvian Council of Science; the Ministry of Education and Science, project no. 2022/WK/14, and the National Science Center, contracts Opus 2021/41/B/ST2/01369 and 2021/43/B/ST2/01552 (Poland); the Funda\c{c}\~ao para a Ci\^encia e a Tecnologia, grant CEECIND/01334/2018 (Portugal); the National Priorities Research Program by Qatar National Research Fund; MCIN/AEI/10.13039/501100011033, ERDF ``a way of making Europe", and the Programa Estatal de Fomento de la Investigaci{\'o}n Cient{\'i}fica y T{\'e}cnica de Excelencia Mar\'{\i}a de Maeztu, grant MDM-2017-0765 and Programa Severo Ochoa del Principado de Asturias (Spain); the Chulalongkorn Academic into Its 2nd Century Project Advancement Project, and the National Science, Research and Innovation Fund via the Program Management Unit for Human Resources \& Institutional Development, Research and Innovation, grant B37G660013 (Thailand); the Kavli Foundation; the Nvidia Corporation; the SuperMicro Corporation; the Welch Foundation, contract C-1845; and the Weston Havens Foundation (USA).
\end{acknowledgments}

\bibliography{auto_generated} 
\cleardoublepage \appendix\section{The CMS Collaboration \label{app:collab}}\begin{sloppypar}\hyphenpenalty=5000\widowpenalty=500\clubpenalty=5000\input{EXO-22-020-public-authorlist.tex}\end{sloppypar}
\end{document}

%% file: EXO-22-020-public-authorlist.tex
\cmsinstitute{Yerevan Physics Institute, Yerevan, Armenia}
{\tolerance=6000
A.~Hayrapetyan, A.~Tumasyan\cmsAuthorMark{1}\cmsorcid{0009-0000-0684-6742}
\par}
\cmsinstitute{Institut f\"{u}r Hochenergiephysik, Vienna, Austria}
{\tolerance=6000
W.~Adam\cmsorcid{0000-0001-9099-4341}, J.W.~Andrejkovic, T.~Bergauer\cmsorcid{0000-0002-5786-0293}, S.~Chatterjee\cmsorcid{0000-0003-2660-0349}, K.~Damanakis\cmsorcid{0000-0001-5389-2872}, M.~Dragicevic\cmsorcid{0000-0003-1967-6783}, P.S.~Hussain\cmsorcid{0000-0002-4825-5278}, M.~Jeitler\cmsAuthorMark{2}\cmsorcid{0000-0002-5141-9560}, N.~Krammer\cmsorcid{0000-0002-0548-0985}, A.~Li\cmsorcid{0000-0002-4547-116X}, D.~Liko\cmsorcid{0000-0002-3380-473X}, I.~Mikulec\cmsorcid{0000-0003-0385-2746}, J.~Schieck\cmsAuthorMark{2}\cmsorcid{0000-0002-1058-8093}, R.~Sch\"{o}fbeck\cmsorcid{0000-0002-2332-8784}, D.~Schwarz\cmsorcid{0000-0002-3821-7331}, M.~Sonawane\cmsorcid{0000-0003-0510-7010}, S.~Templ\cmsorcid{0000-0003-3137-5692}, W.~Waltenberger\cmsorcid{0000-0002-6215-7228}, C.-E.~Wulz\cmsAuthorMark{2}\cmsorcid{0000-0001-9226-5812}
% ...
\par}
\vskip\cmsinstskip
\dag:~Deceased\\
$^{1}$Also at Yerevan State University, Yerevan, Armenia\\
$^{2}$Also at TU Wien, Vienna, Austria\\
% ...
